# Estimation methods for estimands using the treatment policy strategy; a simulation study based on the PIONEER 1 Trial

Authors: James Bell[1], Thomas Drury[2], Tobias Mütze[3], Christian Bressen Pipper[4], Lorenzo Guizzaro[5], Marian Mitroiu[6], Khadija Rerhou Rantell[7], Marcel Wolbers[8], David Wright[9]

**Affiliations**

[1] Clinical Operations, Elderbrook Solutions GmbH, Buckinghamshire, UK

[2] Statistics & Data Science Innovation Hub, GlaxoSmithKline, London UK

[3] Analytics, Novartis Pharma AG, Basel, Switzerland

[4] Methods and Outreach, Biostatistics, Novo Nordisk A/S, Soeborg, Denmark & Dept. of Public Health, University of Southern Denmark, Odense, Denmark

[5] Human Medicines, European Medicines Agency, Domenico Scarlattilaan 6, 1083 HS Amsterdam, The Netherlands

[6] Evidence Generation Biosimilars, Biogen International GmbH, Neuhofstrasse 30, 6340 Baar, Switzerland

[7] Medicines and Healthcare products Regulatory Agency, London, UK

[8] Statistical Methods, Collaboration, and Outreach, Data and Statistical Sciences, Pharma Development, Roche, Basel, Switzerland

[9] Statistical Innovation, Respiratory and Immunology Biometrics and Statistical Innovation, Biopharmaceuticals R&D, AstraZeneca, Cambridge, UK

**Keywords**: estimand, estimation, simulation study, treatment policy, intercurrent event, retrieved dropout, multiple imputation, reference-based imputation.

Abstract (word count: 250)

Estimands using the treatment policy strategy for addressing intercurrent events are common in Phase III clinical trials. One estimation approach for this strategy is retrieved dropout whereby observed data following an intercurrent event are used to multiply impute missing data. However, such methods have had issues with variance inflation and model fitting due to data sparsity.

This paper introduces likelihood-based versions of these approaches, investigating and comparing their statistical properties to the existing retrieved dropout approaches, simpler analysis models and reference-based multiple imputation. We use a simulation based upon the data from the PIONEER 1 Phase III clinical trial in Type II diabetics to present complex and relevant estimation challenges.

The likelihood-based methods display similar statistical properties to their multiple imputation equivalents, but all retrieved dropout approaches suffer from high variance. Retrieved dropout approaches appear less biased than reference-based approaches, resulting in a bias-variance trade-off, but we conclude that the large degree of variance inflation is often more problematic than the bias. Therefore, only the simpler retrieved dropout models appear appropriate as a primary analysis in a clinical trial, and only where it is believed most data following intercurrent events will be observed. The jump-to-reference approach may represent a more promising estimation approach for symptomatic treatments due to its relatively high power and ability to fit in the presence of much missing data, despite its strong assumptions and tendency towards conservative bias.

More research is needed to further develop how to estimate the treatment effect for a treatment policy strategy.

## Introduction

The ICH E9(R1) addendum [1] on estimands and sensitivity analyses in clinical trials provides a framework for the transparent definition of treatment effects of interest that account for intercurrent events (IEs). These are events such as the use of rescue medication, discontinuation of treatment, treatment switching, or death that occur after initiation of allocated treatment, which affect either the interpretation or the existence of the outcome or interest. ICH E9(R1) proposed five strategies for addressing such IEs, and this research focuses specifically on the estimation of treatment effects where the treatment policy strategy is used.

The treatment policy strategy defines a treatment effect that includes the effects of the IEs handled in this way. Upon its introduction, it was widely considered to be similar to the Intention-To-Treat (ITT) principle [1,2]. However, treatment policy is part of a definition of what is being estimated (i.e. the estimand), whereas the ITT principle requires that patients are analysed according to the group they were allocated, and, in stricter definitions [3], that patients are followed up as fully as possible and that all data is included in the analysis regardless of what happens to the patient. This fundamental difference has consequences for estimation [4]: Both cases typically use all available data in estimation but in addition, treatment policy estimation needs to consider how to handle missing data to account for the effects of any previous IEs (i.e. conditioning upon them). In contrast, as ITT does not specify a target of estimation, it is unclear whether the missing data handling of the estimation method should account for the IE (and if so, how). In practice, the occurrence of the IE was generally ignored under ITT estimation [2].

Many of the early approaches to treatment policy estimation continued the existing practices for ITT analysis. Commonly in longitudinal data, a standard Mixed Model Repeated Measures (MMRM) [5] analysis was performed using all available data [6], following the common misconception that a treatment policy strategy meant that IEs are ignored. This approach may be reasonable with a negligible amount of missing data, but as missing data increases, the alignment between the target estimand and the estimation approach diverges for two reasons: Firstly, because standard MMRM cannot distinguish between data before and after an IE, and secondly because missing data are typically highly correlated with the occurrence of IEs [4]. Combined, these can create a strong violation of the missing at random (MAR) assumption required for the validity of inference from the standard MMRM model.

Several sources noted that these simple approaches ignored the potential loss of treatment effect upon treatment discontinuation and made recommendations to adopt methods that had previously been used for conservative sensitivity analysis [6,7,8]. Reference-based imputation (RBI) approaches [9] were commonly advocated, particularly Jump-to-Reference (J2R) methods for treatment discontinuation in placebo-controlled trials with symptomatic treatments [10,11,12] and Copy-Increment-from-Reference (CIR) methods in trials with disease-modifying treatments. These account for the occurrence of IEs when imputing missing data and respectively assume either an instant, complete loss of treatment effect, or a retention of treatment effects accrued until the time of the IE. However, these are both strong assumptions directly about the target of estimation in these patients. They also both make the strong assumption that IEs have no effect in control arm patients. These assumptions have the potential to introduce bias in either direction.

Another set of proposed approaches are the Retrieved Dropout (RD) class. In these, missing data are multiply imputed conditional upon whether they occur pre- or post-IE. Initial attempts used reference sets only containing the appropriate pre- or post-IE data **[13,14]**. However, it proved difficult to fit the imputation models **[14]**, requiring large amounts of post-IE data. Where the imputation models did fit, variance was very high. A simpler approach was then investigated, focusing just on the (typically final) time point of interest **[15]**. This model was usually able to fit, but still suffered from heavily inflated variance and was prone to bias as intermediate visits were discarded. Most recently, research has focussed on all-data imputation models using indicator variables to denote whether subject visits were pre- or post-IE **[16,17,18,19]**. These approaches are more likely to fit than the subsetting models, while also accounting for the intermediate visit data. However, inability to fit and high variance are still a challenge.

Two clear threads running through the research efforts described above are that estimation in the presence of missing data is difficult and that it requires trade-offs between bias and variance.

In this paper we investigate and compare different methods for estimation of treatment effect when a treatment policy strategy is used for handling all IEs, using simulated data that mirrors as closely as possible some of the characteristics of real trial data. We focus on the RD approaches and look at their strengths and weaknesses as the amount of observed off-treatment data increases. We compare them to basic/naïve MMRM and RBI approaches. In doing so, we look at the trade-offs between bias and variance for different methods, when and how often they fail entirely and when and how they can be appropriately used. A presentation of the options, with consolidated advice, is provided to help practitioners navigate an area of statistics where an increasing number of principled methods are being developed, but where balanced comparisons in realistic settings are lacking.

A second aim is to present likelihood-based approaches that try to replicate the RD approaches that have, to date, been implemented by multiple imputation. This would bring practical advantages, such as producing a single fixed estimate, less computational resource usage and removing the need to store many imputed data sets.

The trial we use as the basis for our simulations is the PIONEER 1 study (NCT02906930) **[28]**, which was a 26-week, phase 3a, randomized, double-blind, placebo-controlled, parallel-group trial that assigned 703 patients with equal probability to either placebo or one of three semaglutide doses. The primary objective of the PIONEER 1 study was to compare the efficacy and safety of semaglutide against placebo in patients with type 2 diabetes. The primary endpoint was change in HbA1c from baseline to week 26 and a treatment policy strategy was adopted for all IEs, targeting the effect of oral semaglutide monotherapy versus placebo in the targeted population of patients with type 2 diabetes "regardless of trial product discontinuation or use of rescue medication", i.e., the effect of interest compared treatment strategies that included discontinuation of the randomized treatment and use of rescue medication. It was estimated based on the RD strategy outlined in McEvoy (2016) **[13]**.

This study was chosen firstly as type 2 diabetes was the area that originally motivated many of the regulatory discussions that led to the ICH E9(R1) addendum **[6]**, secondly that it is representative of trials where a treatment policy strategy would be applied to longitudinal continuous data, and thirdly because it is one of the first, and few, published trials with (indirect) information on post-IE outcomes. Our simulation study is of a two-arm trial, so we focus on data from the placebo and semaglutide 3 mg arms. We also replicate the five post-randomisation visits of the original trial.

The paper continues by detailing the statistical methods under investigation in section 2. Section 3 then describes the simulations used to evaluate them. Section 4 provides the results of these simulations, and then section 5 discusses the findings and draws conclusions, including recommendations for which methods to use under which circumstances.

## Methods

### Estimand

For all approaches, the estimand we target is:

*"The difference in mean change from baseline Glycated hemoglobin (HbA1c) at week 26 for type II diabetes patients assigned to either experimental treatment or placebo including any subsequent effects of discontinuation and initiation of rescue medication upon discontinuation."*

More formally, the attributes of the Estimand are as follows:

| Population | Patients with type II diabetes |
| --- | --- |
| Treatment Conditions | Assignment to treatment or control (placebo), including any subsequent discontinuation of the assigned treatment and use of rescue medication. |
| Variable/Endpoint | Change from baseline Glycated hemoglobin (HbA1c) at week 26. |
| Summary Measure | The mean change from baseline Glycated hemoglobin (HbA1c) at Week 26. The between-group comparison is performed using the difference in the mean change. |
| Intercurrent Events (Handling Strategy) | Discontinuation from assigned treatment condition which may also include the use of rescue medication (Treatment Policy). |

### Estimation

This paper considers three general approaches to estimate the estimand defined above in presence of partial post-IE missing data; 1. Estimation via "Simple Models" that make no distinction between outcomes in relation to the IE, 2. Estimation via "Retrieved Dropout Models" that distinguish between outcomes before and after the IE (sometimes referred to as "off-treatment models") and 3. Estimation via "Reference-based imputation" models where a pre-specified reference distribution is used for imputation of post-IE missing data (in our case the placebo group will provide the reference for both groups).

For both the simple and RD models, we consider two different estimation methods. The first uses likelihood-based mixed models for repeated measures **[]** (models denoted with the prefix "MMRM"). In this approach, outcomes at all visits are modelled simultaneously and the treatment effects of interest are estimated directly as the difference between the group marginal means (LS-means) at the week 26 visit. For the RD models, these marginal means also condition on the observed IE proportions at each visit, as explained below. The second estimation method uses sequential multiple imputation (these types of analyses are subsequently denoted with the prefix "MI"). In this approach an MI model is used to create multiple complete datasets and the treatment effects of interest are estimated by fitting an ANCOVA model to each dataset at the week 26 visit. The regression coefficients associated with the treatment effects and standard errors are then combined across multiple imputed datasets via Rubin's rules **[20]**.

For the RBI approaches, we consider imputation under Jump-to-Reference (J2R), Copy-Increments-to-Reference (CIR), and Copy-Reference (CR) assumptions as described in Carpenter et al 2013 **[9]** and handle observed post-IE outcomes using the process described in Wolbers et al 2022 **[21]**.

For the remainder of this section, we introduce notation, describe the different models, and conclude with further discussion points about how statistical inference is performed in each model including how the variance is estimated.

All estimation was performed using SAS, however, it would also be straightforward to implement the MI-based approaches (including the RBI) using R [22] via the "rbmi" package [23]. Example SAS code to illustrate each of the models we discuss is included in the Appendix 1 and full code for the work in this paper is included in a GitHub repository [24].

## Notation

We define $Y_{kij}$ as continuous outcomes for subject $i$ at visit $j$ assigned to group $k$, and $\Delta Y_{kij} = Y_{kij} - Y_{ki0}$ as the corresponding changes from baseline. We let $i = 1, \ldots, n_k$ and $j = 0, \ldots, J$ with zero as baseline and $J$ as the primary visit of interest. We also define two groups with $k \in \{C, T\}$ where $C$ and $T$ correspond to being assigned control and treatment groups, respectively.

We consider a single IE and define $D_{kij}$ to be an indicator variable of IE status, with $D_{kij} = 1$ indicating patient $i$ from group $k$ has experienced an IE **prior to** visit $j$ and $D_{kij} = 0$ otherwise.

For the single IE considered, every patient belongs to one of $J + 1$ IE occurrence patterns: Either they do not experience an IE, or their first visit impacted by the IE is visit $1, \ldots, J$, respectively. We define $P_{kij}$ as a pattern indicator, taking a value of 1 if visit $j$ is the subject's first visit impacted by the IE, and a value of zero otherwise.

Finally, we use Greek letters $\alpha$, $\beta$, $\delta$ for regression coefficients, and $\varepsilon$ for error terms (with suitable indices) for all regression models.

## Simple Models

The simple approaches make no distinction between outcomes with respect to the IE and have no covariate terms relating to the IE in any models. They rely on a missing-at-random (MAR) assumption, where missing outcomes depend only upon observed and modelled covariates as well as previous outcomes. By not modelling the IE, they assume that any missing outcomes would have been similar to the observed data from similar subjects in the same treatment group.

***MMRM1:***

This is a standard MMRM model for the change from baseline, allowing a different mean at each visit per treatment group and adjustment for the baseline outcome at each visit:

$$\Delta Y_{kij} = \alpha_{kj} + \beta_{j0} Y_{ki0} + \varepsilon_{kij} \quad : \quad (\varepsilon_{ki1}, \ldots, \varepsilon_{kiJ}) | Y_{ki0} \sim MVN(0, \Sigma).$$

We fit a single unstructured covariance matrix $\Sigma$ across both groups.

***MI1:***

This approach uses sequential imputation of missing outcomes at visit $j$ ($j = 1, \ldots, J$), respectively, and is based on a Bayesian linear regression model [20] (Rubin, D. B. 1987) depending on the treatment group and past outcomes:

$$Y_{kij} = \alpha_{kj} + \beta_{j0} Y_{ki0} + \beta_{j1} Y_{ki1} + \cdots + \beta_{j(j-1)} Y_{ki(j-1)} + \varepsilon_{kij} \quad : \quad \varepsilon_{kij} | Y_{ki0} \ldots, Y_{ki(j-1)} \sim N(0, \sigma_j^2).$$

## Retrieved Dropout Models

These approaches add flexibility to the simple models by distinguishing between pre- and post-IE outcomes. They are compatible with *extended* MAR assumptions, i.e., they assume that missing outcome data is similar to observed data from subjects in the same treatment group with the same observed outcome history, and either the same IE status (MI2/MMRM2) or pattern (MI3/MMRM3).

### MMRM2:

The MMRM2 model extends the MMRM1 model to allow a two-level distinction between patient outcomes as pre- or post-IE at each visit. This is achieved by including indicators $D_{kij}$ for IE status at each visit per treatment group, resulting in the following model:

$$\Delta Y_{kij} = \alpha_{kj} + \delta_{kj} D_{kij} + \beta_{j0} Y_{ki0} + \varepsilon_{kij} \quad : \quad (\varepsilon_{ki1}, \dots, \varepsilon_{kiJ}) | Y_{ki0}, D_{ki1}, \dots, D_{kiJ} \sim \text{MVN}(0, \Sigma).$$

As with MMRM1, we fit a single unstructured covariance matrix $\Sigma$ across both groups. The treatment effects are estimated as the difference in overall group means, constructed by standardization **[30]** i.e. a linear combination of the pre- and post-IE group means using the observed proportions. A variance correction is also applied to provide model based standard errors that reflect uncertainty in the pre- and post-IE proportions (See Appendix 2 for further details).

### MI2:

MI2 extends the MI1 model using IE status in the same way as MMRM2 extends MMRM1. However, rather than regressing on the observed earlier outcomes as done in MI1, the sequential imputation model regresses on the residuals calculated using predicted values from earlier imputation models. Using the residuals in this way is similar to the residual conditioning in the MMRM approaches and has been shown to reduce bias and variance inflation **[19, 31]**. The regression models used for imputation are given by:

$$Y_{kij} = \alpha_{kj} + \delta_{kj} D_{kij} + \beta_{j0} Y_{ki0} + \beta_{j1}(Y_{ki1} - \hat{Y}_{ki1}) + \cdots + \beta_{j(j-1)}(Y_{ki(j-1)} - \hat{Y}_{ki(j-1)}) + \varepsilon_{kij} :$$
$$\varepsilon_{kij} | D_{ki1}, \dots, D_{kij}, Y_{ki0}, \dots, Y_{ki(j-1)} \sim N(0, \sigma_j^2).$$

where the terms $(Y_{kil} - \hat{Y}_{kil})$ define the residuals using the predicted values $\hat{Y}_{kil}$ from the previous $l - 1$ visits.

### MMRM3:

The MMRM3 model extends MMRM2 beyond the two level pre- and post-IE status to allow the distinction between the timing of IE occurrence. This is achieved by including a set of indicators $P_{ki1}, \dots, P_{kij}$ at visit $j$. As we consider the case where IE occurrence is monotone, this corresponds to the pattern of IE occurrence and results in the following model:

$$\Delta Y_{kij} = \alpha_{kj} + \delta_{kj1} P_{ki1} + \cdots + \delta_{kjj} P_{kij} + \beta_{j0} Y_{ki0} + \varepsilon_{kij} \quad : \quad (\varepsilon_{ki1}, \dots, \varepsilon_{kiJ}) | P_{ki1}, \dots, P_{kiJ}, Y_{ki0} \sim \text{MVN}(0, \Sigma).$$

As with the other MMRM models, we fit a single unstructured covariance matrix $\Sigma$ across both groups. The treatment effects are estimated similarly to MMRM2 but constructed using the means and proportions from the IE occurrence pattern. A variance correction is also applied to provide model based standard errors (See Appendix 2 for further details).

*MI3:*

The MI3 model extends MI2 in the same way as MMRM3 extends MMRM2 resulting in the following model:

$$Y_{kij} = \alpha_{kj} + \delta_{kj1}P_{ki1} + \cdots + \delta_{kjj}P_{kij} + \beta_{j0}Y_{ki0} + \beta_{j1}(Y_{ki1} - \hat{Y}_{ki1}) + \cdots + \beta_{j(j-1)}(Y_{ki(j-1)} - \hat{Y}_{ki(j-1)}) + \varepsilon_{kij}:$$

$$\varepsilon_{kij}|P_{ki1}, \ldots, P_{kij}, Y_{ki0} \ldots, Y_{ki(j-1)} \sim N(0, \sigma_j^2).$$

## Reference Based Imputation Approaches

We implement three different RBI assumptions, which characterize whether the effect of the active drug versus placebo up to the IE is fully maintained ("Copy Increments in Reference" assumption), diminished ("Copy Reference" assumption) or eliminated ("Jump to reference" assumption) after the IE **[8,9]**.

Our implementation follows the specifications in Carpenter et al 2013 **[9]**. However, as we also have observed post-IE outcomes, we modify our reference-based approach to mirror Wolbers et al 2022 **[21]**. In brief, the imputation model is a Bayesian MMRM model fitted to the pre-IE outcomes only. The MMRM allows for different means at each visit per treatment group and is adjusted at each visit for the baseline outcome. Traditionally the model would also have a separate covariance matrix for each treatment group, but we fit a single covariance matrix across both treatment groups to match the structure of the other methods we consider. Based on this imputation model, the chosen RBI assumption, the patient's assigned treatment group, and the timing of their IE occurrence, we derive the mean and covariance matrix of the patient's marginal imputation distribution across all visits as described in Carpenter et al. 2013. Before the imputation of missing outcomes, all observed post-IE outcomes are added back into the data. Imputation of the patient's missing outcomes is then based on sampling from the conditional imputation distribution which is obtained by conditioning their marginal distribution on the observed outcomes.

## RD Pattern Collapsing Algorithm

The RD models specified above include covariates (indicator variables) that are defined based on the occurrence of IEs. Compared to the simple models, they add a substantial number of parameters and may suffer from model fitting issues due to data sparsity. For example, this may happen if all subjects in a particular IE pattern withdraw from the trial before the primary visit of interest. To ensure these models can be fitted in all cases, we propose a pre-specified simplification algorithm to collapse problem cases in a structured way. Full details of the algorithm are provided in the Appendix 3, but we give a general outline of the process below.

For any model fitting problems for the IE status/pattern, we combine it with the nearest IE status/pattern, prioritising the one reflecting an IE occurrence at the previous visit over the next visit (as using future knowledge is less optimal). This introduces an assumption of comparability between the patients in the merged patterns. This algorithm allows the model structure to simplify by merging patterns of IE occurrence and therefore collapsing the problematic patterns until patient outcomes for these simpler groupings exist at each visit, allowing the model to be fitted. For cases with multiple fitting issues, repeated merging of patterns is performed which allows the full IE pattern models (MMRM3 and MI3) to collapse eventually to the two-pattern IE status models (MMRM2 and MI2) and finally, in extreme cases, to the single pattern simple models that make no IE distinction (MMRM1 and MI1).

### Treatment effect estimation and inference

For the MMRM1 model, the treatment effect estimate is the difference in the LS-means between the two treatment groups at the last visit. For the RD MMRM models, we predict the LS-means for each treatment group for IE status (MMRM2) or IE pattern (MMRM3) and construct an overall group mean by weighting these by the observed proportion for each status or pattern respectively. The treatment effect is then defined as the difference of these weighted means between treatment groups. The corresponding standard error is derived based on the delta method (see Appendix 2 for details). Importantly, these derivations rely on the simplifying assumptions that IEs occur independently of the observed outcomes and that the probability of the different IE status and IE patterns follow binomial and multinomial distributions respectively. For all MMRM approaches, hypothesis testing used Kenward-Roger adjustments **[25]**.

The MI-based models (including RBI methods) separate the missing data imputation from the analysis step of the completed data. The analysis model of the completed data is a standard ANCOVA model for the change from baseline at the primary visit of interest with terms for treatment group and baseline. The treatment effect estimate for each completed dataset is the difference in treatment group means. Inference for the overall treatment effect is obtained by pooling the treatment effect estimator (and standard errors) across imputed dataset using Rubin's rules. Using Rubin's rules for the RBI approaches leads to "information anchored" estimates of the variance **[26]**, however, we acknowledge there is an ongoing debate regarding the most appropriate variance estimation for RBI models **[27]**.

## Simulation Setup

### Introducing the PIONEER 1 study

We motivate the setup of and the selection of parameters for the simulation study through the design and the results of the PIONEER 1 study. It addressed its main study objectives by specifying two different estimands: a "treatment policy estimand" and a "trial product estimand" **[29]**. The primary treatment policy estimand was as described in the introduction. The additional "trial product" estimand used a hypothetical strategy to target the treatment effect of oral semaglutide monotherapy versus placebo in the targeted population of patients with type 2 diabetes had all patients remained on their assigned trial product and had rescue medication not been made available. This hypothetical estimand was estimated using an MMRM model that used only data collected prior to the occurrence of the intercurrent events.

Tables A2 and A3 in Appendix 4 Simulation Study show summary statistics of the on-study and on-treatment HbA1c in the PIONEER 1 study. These tables show that 178 patients were randomized to the placebo arm and 175 were randomized to the semaglutide 3 mg arm. Also, in the placebo arm, the HbA1c at the final visit (W26) is missing for 10 patients and that up to the time of final visit, 45 patients either discontinued their trial product or took rescue medication. In the semaglutide 3 mg arm, 8 patients had a missing HbA1c value at the final visit and 26 patients either discontinued their trial product or took rescue medication prior to the final visit. In the simulation study, data will be generated that mimics the results of the PIONEER 1 study.

The estimands in the PIONEER 1 study discuss strategies for two intercurrent events: trial product discontinuation and use of rescue medication. In our simulation study, for simplicity and illustration purposes, we focus on the setting of a single intercurrent event (which can equivalently be considered as two intercurrent events that invariable co-occur). Considering the PIONEER 1 study as the motivating example, the single intercurrent event in the simulation study can be interpreted as patients discontinuing their trial product and initiating rescue medication at the same time.

### Simulating patient-level data

The details and statistical models for the simulation of patient-level data are provided in Appendix 4. Here, we provided a summary. We simulate patient-level data for a two-arm randomized clinical trial with an experimental treatment arm $T$ and a control arm $C$. Per arm, data for 200 subjects are simulated. Our simulation of patient-level clinical trial data comprises four steps:

1. Simulate the patient-level HbA1c per arm under the hypothetical setting that all patients remain on their initially assigned trial product. We assume a multivariate normal distribution for the HbA1c across the visits (weeks 0, 4, 8, 14, 20, 26) with mean and covariance matrix differing between arms. The means and covariance matrix are listed in Table A4 in Appendix 4.
2. Simulate the occurrence of the intercurrent event for each patient at a particular visit. The conditional probability for a patient to have an intercurrent event at visit $j$ without having an intercurrent event at visit $j-1$ is defined through a logistic regression model. We consider two models for treatment discontinuation, which we name following the missing data framework language [20]. In the first model, which we denote as the 'discontinuation at random' (DAR) model, the conditional intercurrent event probability for visit $j$ depends on the study arm, the visit, the baseline HbA1c, and the HbA1c of the preceding visit $j-1$. In the second model, which we denote as the 'discontinuation not at random' (DNAR) model, the conditional intercurrent event probability a visit $j$ depends on the same parameters as in the DAR model and additionally also depends on the HbA1c at visit $j$. The specifics of the DAR model and the DNAR model are provided in Tables A5 and A6 in Appendix 4.

3. Simulate the off-trial-product HbA1c, i.e., simulate patient's trajectory after the occurrence of the intercurrent event. Thereto, if a patient had an intercurrent event, the HbA1c simulated under 1. is shifted by $\delta(\cdot)$. We distinguish two models for the shift $\delta(\cdot)$. In the first model, denoted "Instant change", $\delta(\cdot)$ is a constant HbA1c decrease. In the second model, denoted "Gradual change", $\delta(\cdot)$ corresponds to HbA1c decreases over up to three visits. A decrease in HbA1c after the intercurrent event is aligned with the assumption in our simulation study that the intercurrent events corresponds to patients simultaneously discontinuing the trial product and taking recue medication. The exact shifts are provided in Table A7 in Appendix 4.

4. The missingness indicators are generated and the missing values removed. We also assume that missing observations can only occur for those patients who already had the intercurrent event and that missing is not intermittent, that is once a patient has a missing observation, the observations from all future visits will also be missing. The conditional probability for a patient, who had discontinued their treatment and has not yet had any missing observations, to have a missing observation at visit $j$ is varied between $\text{logit}^{-1}(0.1)$ and $\text{logit}^{-1}(0.6)$ in steps of 0.1 on the logit-scale. Here, $\text{logit}^{-1}(x) = 1/(1 + \exp(-x))$. This defines six missing scenarios which are characterized by an increasing proportion of missing values. The details are provided in Table A8 in Appendix 4.

The properties of the simulated data are closely mimicking both the on-treatment and on-study means of the PIONEER 1 data as illustrated by Figures 1 and 2 in Appendix 4. It is worth emphasizing that the parameter choices for the simulation model are informed based on the summary statistics shown in Tables 2 and 3 in Appendix 4. Thus, a limitation of this simulation setup is that the simulation model might not fully reflect the association between the different events in patient's journey.

**Comparison of the missingness scenarios**

Table *1* and Table *2* show the treatment discontinuation and amount of observation that occurs in each of the missingness scenarios. Table *1* shows the average percentage of patients at the final visit that are

1. still receiving assigned treatment and so observed,
2. discontinuing from assigned treatment but remaining in the trial and therefore observed, or
3. discontinuing treatment but also withdrawing from the trial and therefore missing.

Table *2* presents the percentage of the patients who discontinued that have data observed or missing at the final timepoint. All missingness scenarios have on average 74.5% discontinuations in the control group and 84.8% discontinuations in the Treatment group. Missingness scenario 1 is a setting where substantial amounts of post-IE data

are collected with approximately 70-75% of discontinued patients continuing to provide data (and approximately 20-25% withdrawing), closely reflecting the PIONEER 1 trial data. Missingness scenarios 2 and 3 show a more moderate amount of post-IE data collected with approximately 35-55% of discontinued patients continuing to provide data (and approximately 45-65% withdrawing). Finally, missingness scenarios 4, 5 and 6 are settings where low amounts of post-IE data are collected with approximately 10-30% of patients continuing to provide data (and approximately 70-90% withdrawing).

Table 1 The average percentage of patient status at Visit 5.

| Group | Status (%) | Missingness Scenario | | | | | |
|---|---|---|---|---|---|---|---|
| | | 1 | 2 | 3 | 4 | 5 | 6 |
| Control | Receiving assigned treatment (all observed) | 74.5 | 74.5 | 74.5 | 74.5 | 74.5 | 74.5 |
| | Discontinued assigned treatment and observed | 19.3 | 14.4 | 10.6 | 7.6 | 5.3 | 3.5 |
| | Discontinued assigned treatment and missing | 6.2 | 11.1 | 14.9 | 17.9 | 20.2 | 22.0 |
| Treatment | Receiving assigned treatment (all observed) | 84.8 | 84.8 | 84.8 | 84.8 | 84.8 | 84.8 |
| | Discontinued assigned treatment and observed | 10.8 | 7.5 | 5.1 | 3.4 | 2.2 | 1.3 |
| | Discontinued assigned treatment and missing | 4.4 | 7.7 | 10.1 | 11.8 | 13.0 | 13.9 |

Table 2 Average percentage of patients who discontinued treatment with data observed or missing at Visit 5.

| Group | Discontinued Status (%) | Missingness Scenario | | | | | |
|---|---|---|---|---|---|---|---|
| | | 1 | 2 | 3 | 4 | 5 | 6 |
| Control | Observed | 75.7 | 56.6 | 41.6 | 29.9 | 20.9 | 13.9 |
| | Missing | 24.3 | 43.4 | 58.4 | 70.1 | 79.1 | 86.1 |
| Treatment | Observed | 71.1 | 49.5 | 33.7 | 22.2 | 14.3 | 8.7 |
| | Missing | 28.9 | 50.5 | 66.3 | 77.8 | 85.7 | 91.3 |

## Operating characteristics of interest

The operating characteristics of interest in the simulation study are the bias, standard deviation of estimates, estimated standard error, power, type I error and 95% Confidence Interval (CI) coverage for the treatment policy estimand, that is an estimator for the difference between treatment and control in mean change of HbA1c from baseline to week 26 including any effects of the intercurrent event. The bias and coverage depend on the *true* estimand value. To calculate the true estimand value, a large data set ($n = 200\,000$) with off-trial-product data and without missing data is simulated. For each patient in this data set, the HbA1c change from baseline to week 26 is calculated and the difference in the mean change between groups is considered as the *true* estimand value.

To estimate the operating characteristics, $M = 5000$ trials with individual patient data are simulated based on the statistical model outlined above and with the parameter choices listed in Appendix 4. Each of the data sets is analyzed with the different approaches described in the methods section.

With $m = 1, \ldots, M$ the index of the simulated data sets, $\hat{\delta}_m$ the point estimate, and $\delta$ the *true* estimand value, the operating characteristics are estimated as follows:

$$\widehat{Bias} = \frac{1}{M} \sum_{m=1}^{M} (\hat{\delta}_m - \delta)$$

$$\widehat{SD} = \sqrt{\frac{1}{M-1} \sum_{m=1}^{M} \left(\hat{\delta}_m - \bar{\hat{\delta}}_.\right)^2}$$

Our sample size of 200 per arm is overpowered to detect the observed difference of -0.6% HbA1c from the PIONEER study. PIONEER 1 was powered for the much smaller, but clinically meaningful, difference of -0.45% HbA1c. However, 200 patients per arm was considered the minimum necessary to be representative of late-phase continuous data trials and to avoid small-sample effects. Consequently, to compare the power loss due to missingness for the different estimation approaches in an interpretable way, we consider a "super superiority" (superiority by margin) **[32]** setting of assessing if the effect is at least -0.25% HbA1c. This value was chosen arbitrarily to fix a power of 90% using the ANCOVA analysis performed on the full data (without missingness) for the DAR, "Instant change" scenario. Type I error was calculated under a null scenario against an effect size of 0% HbA1c by a separate set of simulations where both arms are simulated from the control arm parameters.

## Results

We include a selection of tables and figures in the main paper to summarise key results and provide further tables and figures in the supporting information (SI).

Comparing the results for the two IE mechanisms (DAR and DNAR), it appears the "at random" mechanism results in slightly poorer performance in terms of bias and other characteristics across the methods (this may be a feature of the specific data-generating model in our setting) but the difference is small. We therefore present here only the results from the DAR mechanism and the DNAR results may be found in Figures SI1-SI4.

### Model fitting.

There were no model fitting issues relating to data sparsity for the simple approaches (MMRM1/MI1) or the RBI approaches (J2R, CIR, CR). As anticipated, there were data sparsity issues for some simulated trials when fitting the RD approaches (MMRM2/MI2 and MMRM3/MI3) which required the application of our simplification algorithm enable fitting. Table 3 shows the percentage of simulations with DAR mechanism where simplified models were required to make the RD approaches fit (the numbers are identical for both "Instant change" and "Gradual change" scenarios and corresponding MMRM and MI models). For the two-pattern RD approaches (MMRM2/MI2), there were either zero or very few simplifications required for missingness scenarios 1 to 5, but this increased for missingness scenario 6. For the full-pattern RD approaches (MMRM3/MI3), simplifications were needed in all missingness scenarios. Many simplifications were required for the higher missingness scenarios, resulting in a small or zero percentage of cases where the intended model could be fitted. Similar percentages were seen for the DNAR mechanism (Table SI1).

Table 3 Percentage of RD models fitted after simplification ensure fitting.

| Analyses Type | Model Structure | Missing Scenarios | | | | | |
|---|---|---|---|---|---|---|---|
| | | 1 | 2 | 3 | 4 | 5 | 6 |
| MMRM2/MI2 Approaches | 2-pattern model | 100.0 | 100.0 | 100.0 | 99.3 | 96.1 | 85.4 |
| | 1-pattern model | 0.0 | 0.0 | 0.0 | 0.7 | 3.9 | 14.6 |
| MMRM3/MI3 Approaches | 6-pattern model | 81.5 | 56.4 | 20.7 | 3.1 | 0.2 | 0.0 |
| | 5-pattern model | 17.5 | 34.8 | 43.1 | 21.9 | 3.1 | 0.1 |
| | 4-pattern model | 1.0 | 8.1 | 28.6 | 41.4 | 26.0 | 6.5 |
| | 3-pattern model | 0.0 | 0.7 | 6.9 | 26.6 | 44.1 | 35.1 |
| | 2-pattern model | 0.0 | 0.0 | 0.7 | 6.3 | 22.7 | 43.6 |
| | 1-pattern model | 0.0 | 0.0 | 0.0 | 0.7 | 3.9 | 14.6 |

Note: The 2-pattern models in both approaches have the same structure.

The results presented consider the estimation performance of the two- and full-pattern RD approaches using all simulated trials regardless of the simplifications required (assessing a more realistic "analysis process" where encountering sparsity problems in a real trial would require pattern collapsing to fit these models).

### Simple Approaches.

The MMRM1 and MI1 estimation approaches effectively showed the same performance characteristics. They displayed anti-conservative bias for all scenarios, that increased to substantial levels of bias across the larger missingness scenarios. The minimum and maximum biases observed were approximately 0.01% and 0.09% HbA1c respectively. The maximum bias corresponds to 20% of the effect size assumed in the original power calculation of the 3mg semaglutide arm in PIONEER 1 (0.45% HbA1c). (Figure 1).

There were only small increases for the SD of the treatment effects (Figure 2) and average model SEs (Figure 3) as missingness increased, consistent with greater uncertainty in what has been estimated. The minimum and maximum SDs were 0.113 and 0.120 respectively. The model SEs matched the SD of the treatment estimates well. The CI coverage was substantially below 95%, worsening across the increased missingness scenarios, because of the bias induced by fitting these models (Figures SI7-SI10 - Zipper Plots). The lowest coverage was 88%, occurring in the "Gradual change" missingness scenario 6. The type 1 error rate for these models was also preserved (Table 4).

The power for MMRM1 and MI1 increased across the missingness scenarios from 91 to 95% resulting from the anti-conservative bias outweighing the small increases in average model SEs (Figure 4).

### RD Approaches.

The MMRM2 and MI2 approaches showed similar properties to each other. Both were unbiased for the "Instant change" settings except in missingness scenario 6 where some anti-conservative bias was observed in similar amounts in each approach. This bias derives from the need to simplify the model to the biased single pattern model (MMRM1 or MI1). In the "Gradual change" settings, both approaches showed increasing anti-conservative bias as the missingness scenarios increased. The bias seen in these settings is driven by the mis-specified two-pattern IE structure which only distinguishes between pre- and post-IE outcomes (allowing adjustment for an average post-IE difference across all IE occurrence patterns). The maximum bias observed was approximately -0.05 % HbA1c, corresponding to 11% of PIONEER 1's planning assumption (Figure 1). In all cases, the bias was considerably smaller than with the simple approaches.

The SDs of the treatment effect estimates for the MMRM2 and MI2 approaches were larger than the corresponding simple analyses (MMRM2/MI2 vs MMRM1/MI1 respectively). For both "Instant change" and "Gradual change" settings, the SD inflation grew rapidly across missingness scenarios 1 to 6 (Figure 2). The minimum and maximum SDs were 0.114 and 0.160 respectively (occurring in missingness scenarios 1 and 6), corresponding to increases of <1% and 33% compared to those of the simple approaches. For missingness scenarios 5 and 6, the SD inflation is an underestimate of the impact of the actual MMRM2 or MI2 models as some of the most data sparse simulations needed simplification to MMRM1 and MI1.

The average model SEs for the MI2 approach mimicked the standard deviations of the point estimates closely and resulted in approximately correct coverage for lower missingness scenarios. However, for the MMRM2 analysis, the average SEs were slightly higher, with larger deviations in the lower missingness scenarios (Figure 3). These larger SE estimates for the lower missingness scenarios resulted in slight over-coverage (Figures SI7-SI10 - Zipper Plots). In general, the coverage for both approaches reduced across missingness scenarios 1 to 6 due to bias induced from pattern simplification. Neither the MMRM2 nor the MI2 approach showed any increase in type 1 error rate (Table 4).

The power for both MMRM2 and MI2 declined rapidly across missing scenarios. The maximum and minimum power occurred in the DAR "Instant change" setting with approximately 88% in scenario 1 reducing to 65% in scenario 6

(Figure 4). The reduction in power is primarily the result of the increase in uncertainty of the post-IE parameter estimates.

Table 4: Table showing the type 1 error rate associated with each method under the "sharp" null of zero effect and equal (DAR or DNAR) IE rates in each group.

|                   | Missing Scenarios |      |      |      |      |      |
| Estimation Method | 1    | 2    | 3    | 4    | 5    | 6    |
| --- | --- | --- | --- | --- | --- | --- |
| FULL  | 4.93 | 4.93 | 4.93 | 4.93 | 4.93 | 4.93 |
| MMRM1 | 4.74 | 4.92 | 4.74 | 4.61 | 4.61 | 4.77 |
| MMRM2 | 4.55 | 5.13 | 5.19 | 5.04 | 4.67 | 4.98 |
| MMRM3 | 4.38 | 4.61 | 4.50 | 4.83 | 4.83 | 5.28 |
| MI1   | 4.78 | 4.84 | 4.76 | 4.67 | 4.50 | 4.75 |
| MI2   | 4.92 | 5.44 | 5.33 | 5.23 | 4.93 | 5.12 |
| MI3   | 4.67 | 4.84 | 4.73 | 4.73 | 4.90 | 5.16 |
| CIR   | 4.00 | 3.32 | 2.72 | 2.36 | 2.12 | 2.06 |
| CR    | 4.08 | 3.06 | 2.44 | 2.02 | 1.66 | 1.62 |
| JTR   | 3.82 | 3.04 | 2.32 | 1.82 | 1.56 | 1.48 |

Note: Simulations are based on 5,000 simulated data sets per scenario, which provide a Monte Carlo standard error for type I error estimates of approximately 0.3%.

The MMRM3 and MI3 approaches also showed similar properties to each other. Both analyses were unbiased for the "Instant change" and "Gradual change" settings except in the extreme missingness scenarios (Scenario 6 only for "Instant change" and 5 and 6 for "Gradual change"). In these extreme missingness cases, multiple pattern simplifications were needed (Table 3). Many models that were equivalent to MMRM2/MI2 and MMRM1/MI1 were needed and only a moderate percentage of cases allowed a more nuanced pattern-based IE distinction. The maximum bias observed was approximately -0.03 % HbA1c, corresponding to 7% of the assumed effect size of -0.45 % HbA1c in PIONEER 1 (Figure 1). In all cases, the bias was smaller than both the simple and two-pattern IE status RD approaches.

The SD of the treatment effects estimates for the MMRM3/MI3 approaches were similar and only diverged slightly in the larger missingness scenarios (Figure 2). Both approaches resulted in larger SDs than the corresponding MMRM2/MI2 approaches and therefore much larger than the simple MMRM1/MI1 approaches. In both the "Instant change" and "Gradual change" settings, the inflation was extreme across the larger missingness scenarios (Figure 2). The minimum and maximum SDs were 0.115 and 0.171 respectively, corresponding to increases of 2% and 43% when compared to the SD achievable with the simple approaches. As with MMRM2/MI2, this would be an underestimate of the models themselves due to the large number of simplifications to lower variance models occurring.

As with the two-pattern IE approaches, the average model SEs for the MI3 approach mimicked the SDs of the estimates closely and resulted in approximately correct coverage for lower missingness scenarios. However, the average model SEs for the MMRM3 approach were a little too high, with larger estimates in the lower missingness scenarios (Figure 3) – again resulting in over coverage (Figures SI7-SI10 - Zipper Plots). In general, the coverage for both approaches reduced across missingness scenarios 1 to 6, again due to bias induced by pattern simplification. Neither the MMRM3 nor the MI3 approaches increased type 1 error (Table 4).

Power for both MMRM3 and MI3 declined even more rapidly across missing scenarios than for the two-pattern IE approaches. The maximum and minimum power occurred in the DAR "Instant change" setting with approximately 87% in scenario 1 reducing to 58% in scenario 6 (Figure 4). Again, the reduction in power is driven by the increase in uncertainty of the post-IE parameter estimates.

### RBI Approaches.

The RBI approaches showed differing results depending on the scenario. For the "Instant change" setting, the J2R analysis was close to unbiased, showing a slight increase in conservative bias as the missingness scenarios increased which then reverted towards the truth in the extreme cases in scenarios 5 and 6 (Figure 1). The CR method was next best, with anti-conservative bias that worsened across missingness scenarios 1 to 6. Finally, the CIR approach was the worst performing, with more extensive anti-conservative bias that worsened with increasing missingness. For the "Gradual change" setting all RBI methods showed bias and the order of performance remained the same. The maximum bias across either post-IE scenario was approximately -0.01, -0.025 and -0.045 for J2R, CR and CIR respectively corresponding to an absolute percentage of 2%, 6% and 10% of the assumed effect size of -0.45 % HbA1c in PIONEER 1 (Figure 1).

The SD of the treatment effects estimates for the RBI approaches were almost identical and constant across all missingness scenarios. In all cases, they resulted in slightly smaller variation than the full data estimates and were constant because no estimation of post-IE behavior occurs (Figure 2).

The average model SEs for all the RBI approaches were also nearly identical and followed a similar pattern to the simple methods (MMRM1/MI1). They were also consistently higher than the SDs of the treatment effect estimates. The minimum and maximum SEs for the RBI methods were approximately 0.114 and 0.124 for missingness scenario 1 and 6 respectively. This corresponds to an approximate increase of 1% and 3% when compared to the simple methods.

All RBI methods resulted in over coverage in both the "Instant change" and "Gradual change" settings for all missingness scenarios except the CIR method in missingness scenarios 4, 5 and 6 where the sizable bias reduced the coverage. (Figures SI7-SI10 - Zipper Plots). None of the RBI methods increased the type 1 error and in general, it was considerably lower than the simple or RD methods (Table 4).

None of the RBI methods had any substantial reduction in power in any of the missing scenarios for either the DAR "Instant change" or "Gradual change" settings (Figure 4). The CIR method resulted in an increase in power as missingness increased, similar to the simple methods that were also anti-conservative. The minimum and maximum powers were approximately 90% and 95% in missing scenarios 1 and 6 respectively. The CR method produced largely constant power of 90% across all missing scenarios. The J2R method showed a slight reduction in power as missingness increased in both the "Instant change" and "Gradual change" scenarios, with maximum power of 90% in missing scenario 1 and minimum power of 85% in missing scenario 6.

## Discussion

This research set out to compare several methods for estimating estimands with a single type of IE that is addressed by a treatment policy strategy. The goal was to assess the performance of three general approaches for estimating treatment effect in an RCT: First, simple approaches that use all observed data and make no distinction between outcomes with respect to IE occurrence. Second, more complex RD methods that use all observed data with flexibility to distinguish between either IE occurrence or IE pattern (implemented by both MI and mixed models). Third, methods that use all observed data but use reference-based MI to handle any missing post-IE data. The methods were assessed using a simulation study designed to mimic the outcomes, intercurrent events and missingness of the Phase 3 RCT PIONEER1 **[28]** as a base case, with more extreme missingness scenarios also created.

The results of this work imply that the use of simple approaches such as MMRM1 and MI1, which do not make a distinction between pre- and post IE outcome data, can lead to sizable bias for the treatment effects employing a treatment policy strategy, even where most post-IE outcomes are successfully collected, as in missingness scenario 1. We strongly recommend not prespecifying these simple approaches for treatment policy strategy estimation unless only a trivial amount of missing data is expected.

The approaches that make a binary distinction between outcomes pre- and post-IE (MMRM2 and MI2) perform well in settings where the post-IE outcomes are expected to undergo an "Instant change" and the observation of post-IE outcomes is around or above 50% at the timepoint of interest for the analyses (missingness scenarios 1 and 2). These models are less suited to situations where patients take more than one visit to fully transition between on- and off-treatment clinical states as the model cannot account for the number of visits a patient has been off-treatment for. A drawback to these models in general is the loss of power relative to the best performing RBI model (J2R), which becomes meaningful for post-IE observation percentages below 50% and increases at an accelerating rate as this percentage decreases.

The pattern approaches (MMRM3 and MI3) that allow distinction between IE occurrence pattern (and therefore time) do not show any bias in either the "Instant change" and "Gradual change" settings for the majority of missingness scenarios we consider. However, the variance inflation that results from the extra parameters in the model provides a serious concern about the achievable precision, with considerable loss of power relative to the J2R model in all but the lowest missingness scenario (1). The models also suffer from fitting problems and may require at least some simplification even for the lowest missingness scenarios. These fitting problems likely prevent the use of even more complex retrieved dropout models to handle different types of intercurrent events separately. In general, it is questionable if the situationally smaller bias than other methods is worth the fitting problems and substantial decrease in precision, unless there is a strong expectation of at least 60-70% post-IE data collection (similar to missingness scenario 1). The sharply decreased power if the required level of data collection does not occur poses a serious risk to successfully quantifying the effect of a treatment policy in a clinical trial.

The RBI approaches show mixed success. In cases where the assumption chosen for the missingness reflects (or partially reflects) the true post-IE behavior, such as J2R in our simulation, the approach performs well. The core J2R assumption (immediately loss of treatment effect due to the IE) is a reasonable fit for the clinical settings investigated, namely a placebo-controlled trial where the treatment effect mostly disappears post-IE and does so quickly.

The CR approach performs reasonably, perhaps as it inherently assumes retention then slow loss of treatment effect so that for many patients the assumed remaining treatment effect had mostly disappeared by the final visit. This approach is perhaps most suitable for clinical situations where loss of treatment effect occurs but only slowly relative to the trial visit schedule. That it has reasonable characteristics despite its strong statistical assumption not matching the clinical setting suggests that RBI methods may still be reasonable in settings where there is some, but not extensive, mismatch.

In contrast, the CIR approach does poorly, with substantial bias in all investigated settings, likely because CIR inherently assumes a preservation of the treatment effect present at the time of discontinuation. CIR would therefore seem inappropriate for symptomatic-relief settings where treatment effect is rapidly lost, but most suitable for disease modifying treatments settings. This highlights that where the post-IE effect assumption chosen is drastically different to the true post-IE behavior, RBI approaches can be just as biased as the simple approaches. The need to identify *a priori* an appropriate mechanism based on the indication and mode of action of the treatment is a difficulty and potential weakness of RBI approaches.

One large advantage of RBI over RD approaches is the consistently smaller SDs of the estimates. This is because RBI derives the uncertainty for the missing data from the whole control arm, whereas the RD approaches derive it from only the much smaller post-IE data within each arm. In our manuscript, the model-based SEs were estimated using Rubin's rules. Due to the so-called "information-anchoring" property of Rubin's variance, these SEs behaved similarly to those for the simple approaches **[26]**. It has been demonstrated that Rubin's variance overestimates the repeated-sampling variance of treatment effect estimators of RBI approaches **[27, 33]** and the model-based SEs for RBI approaches were noticeably larger than corresponding SDs of point estimates in our simulation study. This occurs because RBI methods borrow information from reference group data for imputations in the intervention group. This induces a positive correlation between outcomes in the two treatment groups and a reduction in the variance of the resulting treatment effect contrast, which is not captured by Rubin's variance estimator. Indeed, as illustrated in our simulations, the strong RBI assumptions may lead to SDs which decrease in magnitude with increasing amounts of missing data. As an alternative to Rubin's variance, so-called frequentist variance estimators have been proposed which might more accurately target the actual variability of the estimator and lead to more accurate type I error control, **[21,27]**.

When using RBI methods, the choice of, and belief in, assumed post-IE behavior should be a clinical, not statistical matter, based on factors such as the disease course and the mechanism of action of treatment. It is important to consider these carefully when choosing between J2R, CR and CIR approaches. There is also a potential issue in that the assumptions are not informed by what is observed post-IE, and in some cases, a direct conflict may be observable (or even testable). It is questionable whether it is statistically appropriate to make (very strong) assumptions directly on the value of the quantity you are trying to estimate.

Comparing the average estimates of the MMRM and MI approaches in general shows that they almost exactly match, which is in line with our expectations. The SDs of the point estimates are also very similar, and they align with the SEs derived from the MI approaches. However, the MMRM model-based SEs do not appear to be as accurate. The variance correction we propose appears to be necessary but inadequate, with systematic discrepancies remaining between the SDs of the estimates and the calculated SEs; the SEs are a little too high. This is likely due to the simplifying assumption that the IE proportion and the outcomes are independent, as we simulate poorer outcomes being linked to discontinuation. The unaccounted-for correlations between outcome and probability of IE occurrence are therefore the likely source for the discrepancies. Further work to investigate and correct this, possibly through joint modelling, would be of interest. Despite the slightly lower efficiency, there are potentially some practical benefits of the MMRM approaches over the MI methods, since they reduce computational time and are logistically easier as a basis for subgroup analyses (which may require repeating the MI for each subgroup).

Overall, for trial data similar to PIONEER 1 (corresponding to missingness scenario 1), the MI2, MMRM2, MI3, MMRM3, CR and J2R methods are all reasonable estimators with little to choose between them. However, by missingness scenario 2, corresponding to approximately 50% post-IE retrieval, only really the MI2, MMRM2, CR and J2R methods maintain both low bias and variance. This is important as, from the authors' experience, 50% post-IE retrieval is a realistic expectation for continuous outcomes in many indications if good trial conduct processes are performed. For greater missingness, only the J2R method performs acceptably at delivering both low bias and variance. At these levels, the J2R method also delivers substantially higher power compared to the RD methods. This is relevant as for some, particularly CNS, indications even 50% post-ICE retrieval may not be realistically obtainable. The strong performance of J2R is despite a small beneficial treatment effect remaining after discontinuing in our simulated data.

As all methods preserve type I error, it is hard to argue against the choice of J2R where there is confidence the treatment does not modify the underlying disease, since it is robust to higher-than-expected missingness. MMRM2 or MI2 could be considered where you are confident of strong post-IE observation (>50%), but even here, the benefits over J2R are limited as its properties also improve as more data is collected. Due to the extreme sensitivity of RD methods' power to missing data, where an RD model is pre-specified as a main analysis we also recommend pre-specifying a percentage threshold for post-IE observation below which another method (likely an RBI) would be used instead.

The fundamental problem with treatment policy estimation is the high sensitivity to the extent of missing data that occurs post-IE, which as we have demonstrated, is strongly linked to statistical properties. We repeat the strong advice provided previously **[1,3,4,19]** that when treatment policy based estimands are specified, every effort must be taken to collect all patient data post-IE. However, it should also be recognized that there are practical limitations to this; patient withdrawal of consent, loss to follow-up, difficulty of measurements, enrollment in other clinical trials, patient state of mind and reduction of incentives to undertake procedures after treatment discontinuation can all prevent effective observation of post-IE data. It is therefore important, before specifying a treatment policy based estimand and estimator, to realistically assess the post-IE observation rate if appropriate procedures are performed.

This work is the most comprehensive comparison of simple MAR approaches, RD methods and RBI methods in a realistic simulation setting to date. It is also the first publication to look at MLE approaches for RD models. Previously, Drury et al. [19] assessed models with the same structure as MI2 and MI3 (among others), but with a simulation study where the models were always fittable (post-IE data was guaranteed). Our work creates a more realistic discontinuation set up and shows that although the binary and pattern-based IE approaches can produce unbiased estimates there are many realistic situations where these approaches may not be feasible. Wang et al [15] advocated the use of similar models to MI2 but included only subjects with an IE in the imputation model for missing post-IE data. We would expect this to perform worse in comparison to the MI2 model as there is less information being used but a more formal comparison between these methods would be interesting. Polverjan and Dragalin [17] and Noci et al. [18] also compared similar MAR, RD, and RBI MI-based methods but focused on disease-modifying settings in Alzheimer's and Parkinson's disease respectively.

The first major limitation of this work is that we ignore disease-modifying settings, which prevents our ability to draw conclusions there. The second is we only focus on the one indication, type II diabetes, and it is possible conclusions might vary between indications. However, Drury et al (REF) reached similar conclusions regarding MI2 and MI3 in a respiratory setting with various IE occurrence and missingness rates. The third limitation is we only consider a single (combined) IE, whereas multiple types of IE occur in real trials. We believe that our conclusions remain valid if post-IE observation rates are independent of the IE type, although we recognize this will not always be true. We have also shown difficulties in dealing with one type of event, and it is unlikely that any of the methods investigated would extend well to multiple events: RD approaches already suffer from fitting problems and variance inflation, and these will only worsen with more complicated models. There are no obvious acceptable and differing assumptions that RBI methods could make for different IE types, so they would likely default to not distinguishing between them. The possibility of multiple IEs occurring in a single patient adds further complexity. Each of these situations is important and clinically relevant avenues for future research in treatment policy estimation.

One smaller limitation is that we do not consider intermediate missingness. We do not consider this a major issue as intermediate missingness is typically considered non-informative in clinical trials, and the visit of most interest is usually the last. We also only consider missingness completely at random conditional upon an IE. We believe this is reasonable as most informative factors should be accounted for by the conditioning on the IE. However, this would be a potential future avenue of research.

# Conclusions

Handling IEs *via* treatment policy is easy to specify at the estimand level, but hard to reliably estimate. Simple approaches to deal with missing data, that ignore distinctions between pre- and post-IE data, are biased and not appropriate unless only a trivial amount of missing data is expected. More complex RD models that do condition on IE occurrences appear to address the bias issue but at the expense of substantially decreasing power as the observation of post-IE data decreases. The alternative RBI approaches (in particular, J2R for treatments with no disease-modifying effects) can produce good statistical properties, but only if their strong assumptions are reasonable.

Whenever pre-specifying treatment policy based estimation, a statistician is thus faced with a dilemma; either to make very strong assumptions directly about the magnitude of the treatment effect that could lead to bias if false, or weaker assumptions where insufficient observation of post-IE data will lead to variance inflation and greatly reduced power, potentially undermining the trial's chance of achieving its objectives. For this reason, none of the approaches we looked at are truly satisfactory.

In contrast to the authors' expectations before beginning this study, and somewhat reluctantly given the strong assumptions, we conclude from our simulations that the reference-based J2R is probably the least worst choice from the methods investigated for estimating an Estimand applying the treatment policy strategy to all IEs, in clinical settings with treatments that do not modify the underlying disease ('symptomatic treatments'), particularly where it is not firmly believed that at least 50% of post-IE data will be collected. Its inherent (over-)conservation of type I error may appeal to regulators, while its substantially higher power than RD methods may appeal to sponsors. The MI2/MMRM2 methods also appear useful in these settings if at least 50% post-IE observation is expected. MI3/MMRM3 may be of some use as a sensitivity analysis against deviations from the non-disease-modifying assumption, but they seem too unreliable and variance-inflating to be sensibly pre-specifiable for primary analysis.

As statisticians, we would prefer not to have to make such unenviable choices. We believe it ought to be possible to deal with the variance inflation of RD methods by Bayesian methods incorporating into them a clinically plausible prior belief of, e.g., no treatment effect post-IE, so that with a lack of data the estimator becomes similar to a *de facto* RBI method. This would have the added advantage that the arbitrariness of the RBI variance is now directly relatable to the strength of the prior. We welcome the work of Cro et al. **[34]** in this general direction. We believe such a new class of methods could combine the best of both RD and RBI approaches and that this could be the future of treatment policy estimation.

## Acknowledgement
The authors would like to thank Helle Lynggaard for providing additional precision to the summary information about PIONEER 1, Ian R. White for general discussions on MI for treatment policy estimation and James H. Roger for discussion and algebra relating to MLE with time dependent covariates.

## Data availability statement
All data used in this paper is simulated. All simulation and analysis code is freely available on Github **[24]**.

Figure 1: Bias in the estimated treatment effects at the final timepoint for the DAR mechanism across different post-IE and missingness scenarios

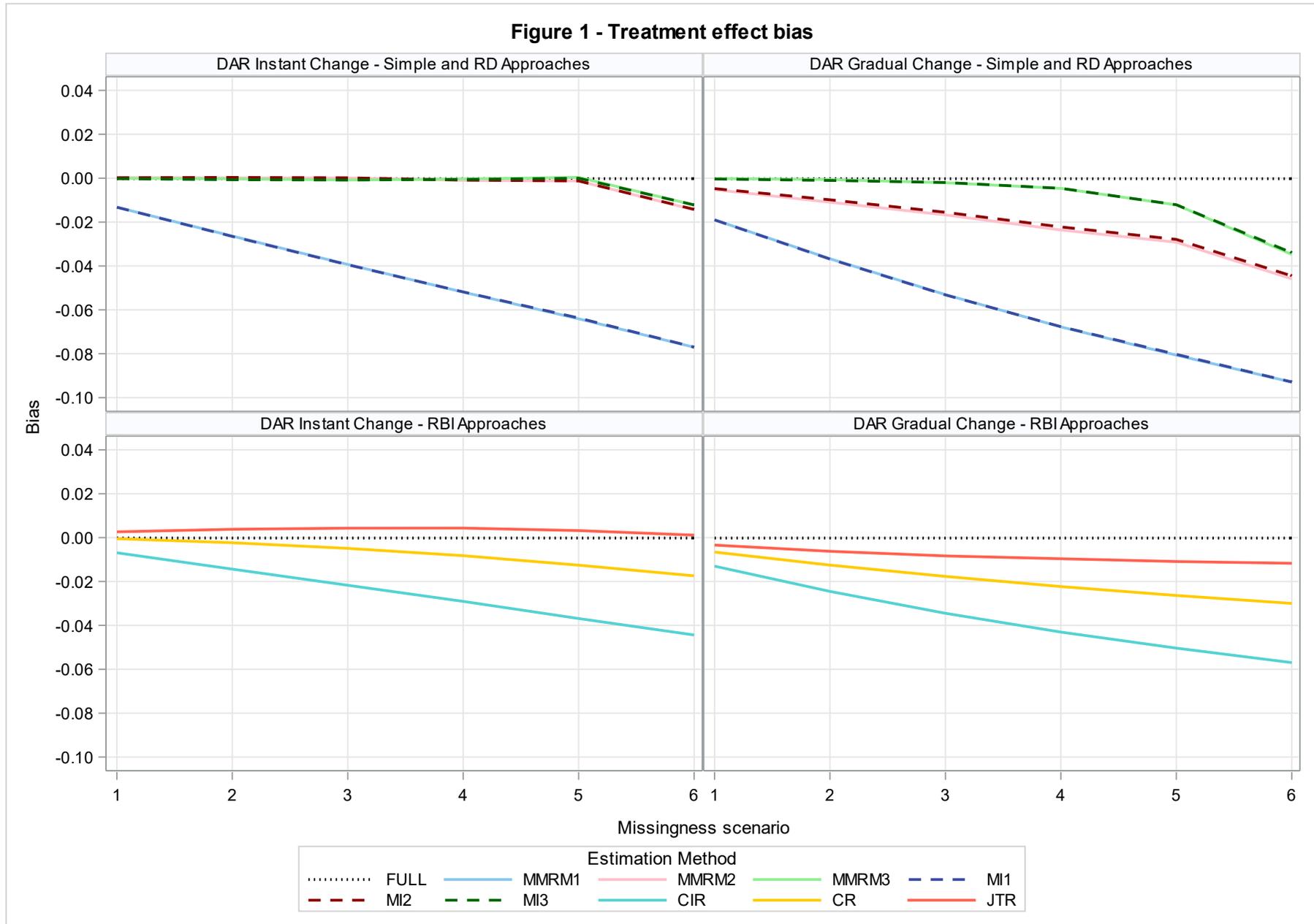

Figure 2: SD of the estimated treatment effects at the final timepoint for the DAR mechanism across different post-IE and missingness scenarios

**Figure 2 - SD of treatement effect estimates**

Figure 3: Average model SE of the estimated treatment effects at the final timepoint for the DAR mechanism across different post-IE and missingness scenarios.

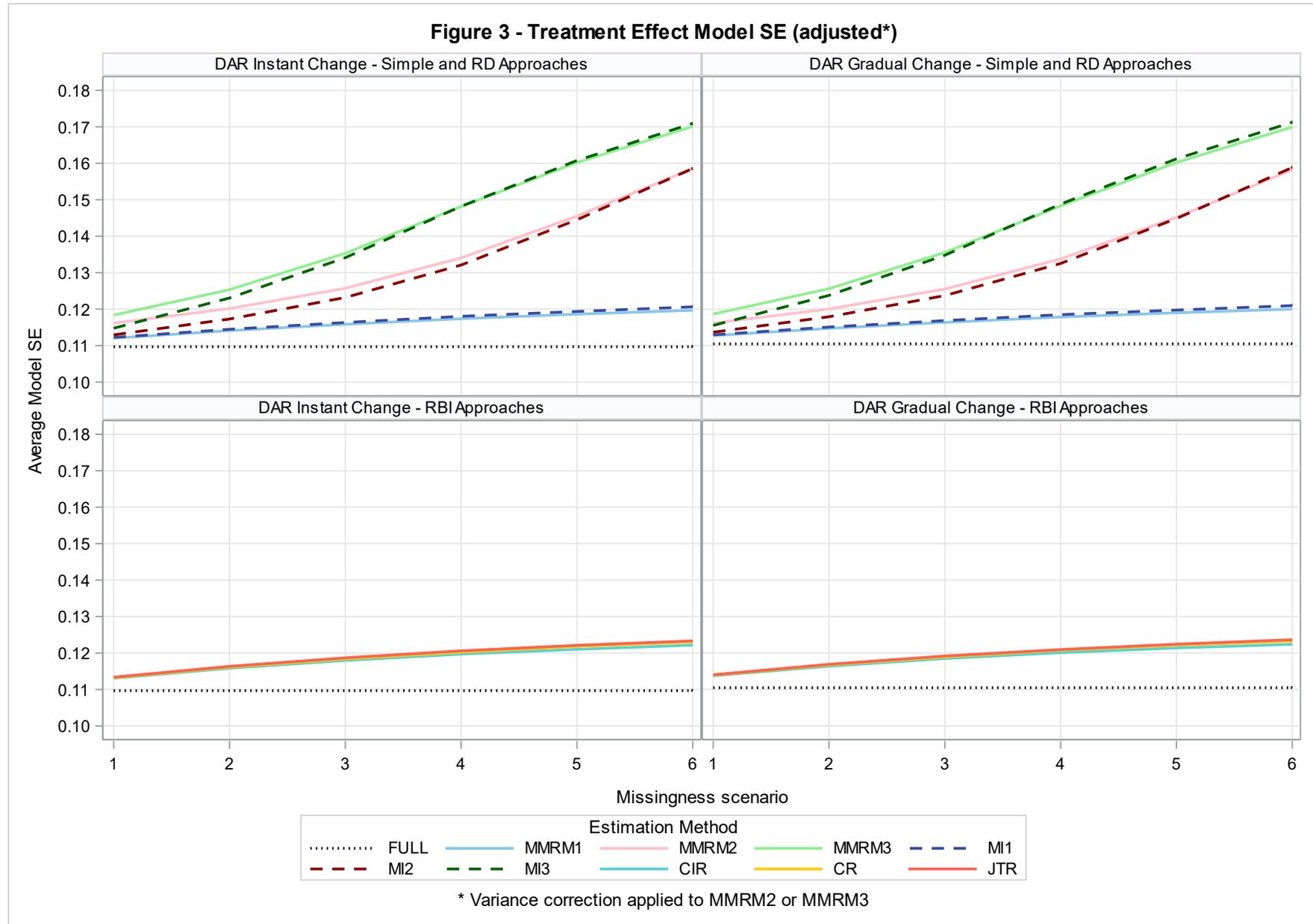

Figure 4: Treatment Effect Super Superiority Power

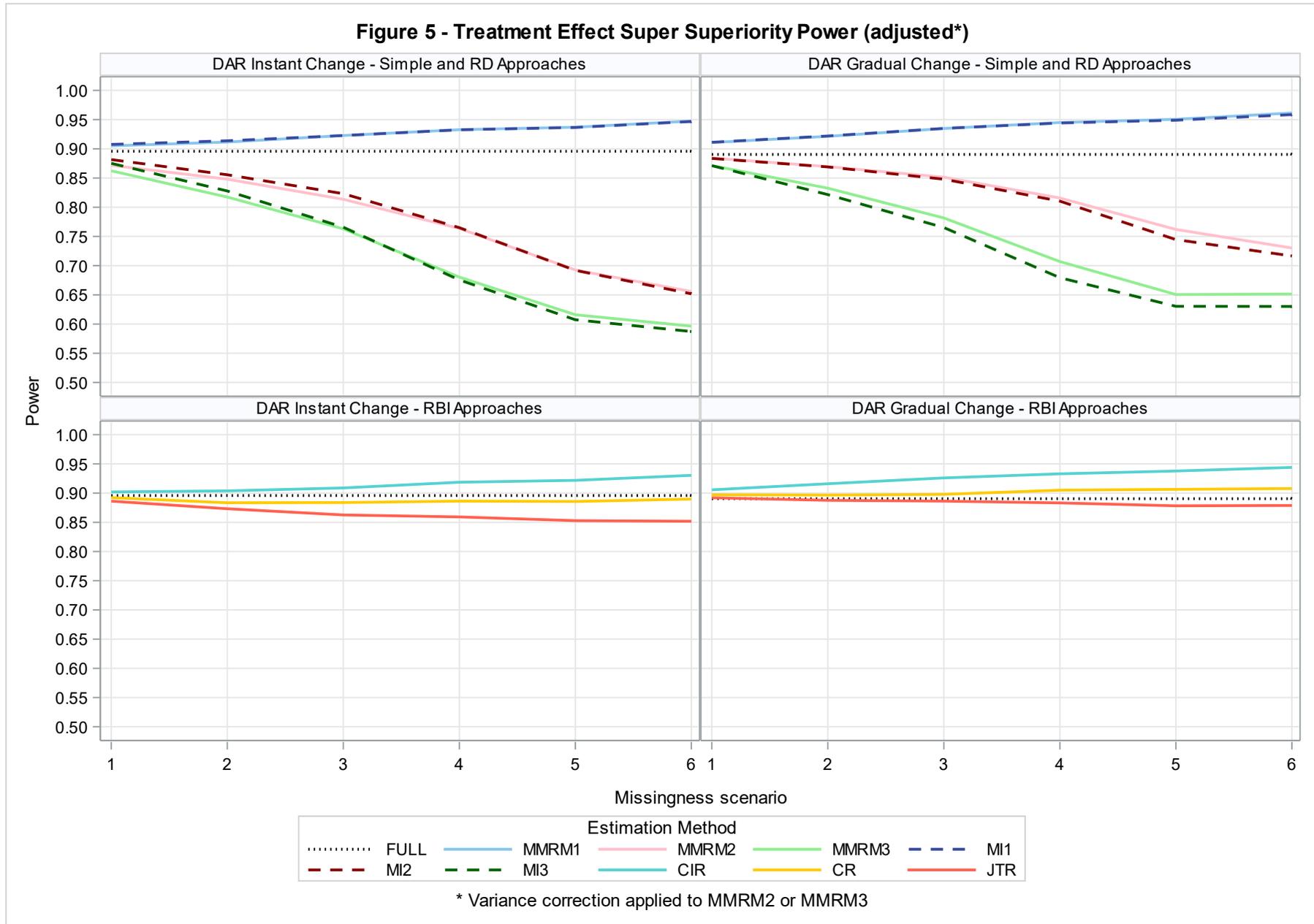

Figure 5: Treatment Effect 95% CI Coverage

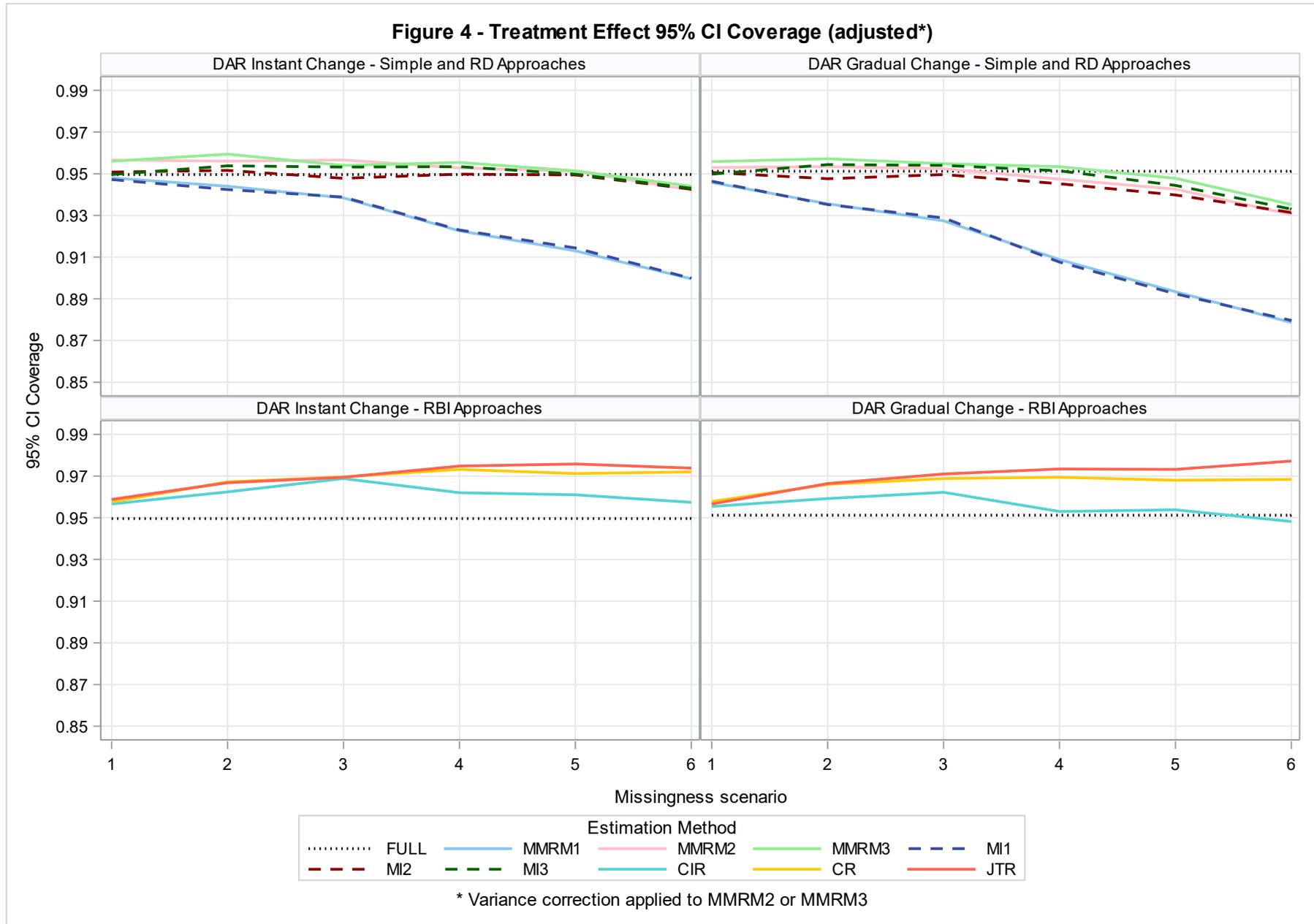

# Appendix 1: Example Model Code

**Structure of MI1**

```
proc mi data = ds out = mi1 nimpute = 50 seed = 12345;
  class TRT;
  var TRT BASE Y1-Y5;
  monotone reg (Y1 = TRT BASE);
  monotone reg (Y2 = TRT BASE Y1);
  monotone reg (Y3 = TRT BASE Y1 Y2);
  monotone reg (Y4 = TRT BASE Y1 Y2 Y3);
  monotone reg (Y5 = TRT BASE Y1 Y2 Y3 Y4);
run;
```

**Structure of MI2**

MI2 model is fitted by regressing on the residuals. To achieve this in base SAS code, we use PROC MI and PROC MIXED sequentially to obtain the residual after each imputation. As baseline is observed for all patients and there can be no IEs, it is added directly without the need for residuals. The code below is an illustration of how each stage of the sequential imputation would proceed. Code in red brackets indicate what would need to change at each stage of the sequential imputation. Full code for these models is provided in the GitHub repository **[24]**.

```
proc mi data = ds out = mi2 nimpute = 50 seed = 12345;
  class TRT D<j>;
  var TRT D<j> BASE R1-R<j>;
  monotone reg (Y<j> = TRT D<j> TRT*D<j> BASE R1 ... R<j-1>);
run;

proc mixed data = mi2;
  class TRT D<j>;
  model Y<j> = TRT D<j> TRT*D<j> BASE Y1 ... Y<j-1> / outpm=mi2_r<j>;
run;
```

**Structure of MI3**

```
proc mi data = ds out = mi3 nimpute = 50 seed = 12345;
  class TRT P1 P2 P3 P4 P5;
  var TRT P1 P2 P3 P4 P5 BASE Y1-Y5;
  monotone reg (Y1 = TRT P1 TRT*P1 BASE);
  monotone reg (Y2 = TRT P2 TRT*P2 BASE Y1);
  monotone reg (Y3 = TRT P3 TRT*P3 BASE Y1 Y2);
  monotone reg (Y4 = TRT P4 TRT*P4 BASE Y1 Y2 Y3);
  monotone reg (Y5 = TRT P5 TRT*P5 BASE Y1 Y2 Y3 Y4);
run;
```

### Structure of MMRM1

```
proc mixed data = ds;
  class VISIT TREATMENT SUBJID;
  model CHANGE = VISIT*TREATMENT VISIT*BASELINE / noint ddfm=kr;
  repeated VISIT / patient=SUBJID type=un;
  lsmeans VISIT*TREATMENT;
run;
```

### Structure of MMRM2

MMRM2 is an extension of MMRM1 with the addition of a time dependent covariate that indicates the **current status** of the patient with respect to IE occurrence.

```
proc mixed data = ds;
  class VISIT TREATMENT DISC SUBJID;
  model CHANGE = VISIT*TREATMENT*DISC VISIT*BASELINE / noint ddfm=kr;
  repeated VISIT / patient=SUBJID type=un;
  lsmeans VISIT*TREATMENT*DISC / cov;
run;
```

Table below illustrates the levels of the indicator DISC in the MMRM2 model in the PIONEER Simulation.

| Pattern (B = Baseline: O = No IE: X = IE) | Description | Disc. Indicator Codes | | | | |
|---|---|---|---|---|---|---|
| | | V1 | V2 | V3 | V4 | V5 |
| BXXXXX | IE after Visit 0 (Post IE at Visit 1) | 1 | 1 | 1 | 1 | 1 |
| BOXXXX | IE after Visit 1 (Post IE at Visit 2) | 2 | 1 | 1 | 1 | 1 |
| BOOXXX | IE after Visit 2 (Post IE at Visit 3) | 2 | 2 | 1 | 1 | 1 |
| BOOOXX | IE after Visit 3 (Post IE at Visit 4) | 2 | 2 | 2 | 1 | 1 |
| BOOOOX | IE after Visit 4 (Post IE at Visit 5) | 2 | 2 | 2 | 2 | 1 |
| BOOOOO | IE does not occur (Completers) | 2 | 2 | 2 | 2 | 2 |
| | Total Covariate Levels | 2 | 2 | 2 | 2 | 2 |

### Structure of MMRM3

MMRM3 is a further extension of MMRM2 with the addition of a time dependent covariate that indicates the **current status** of the patient with respect to their IE pattern.

```
proc mixed data = ds;
  class VISIT TREATMENT PATTERN SUBJID;
  model CHANGE = VISIT*TREATMENT*PATTERN VISIT*BASELINE / noint ddfm=kr;
  repeated VISIT / patient=SUBJID type=un;
  lsmeans VISIT*TREATMENT*PATTERN / cov;
run;
```

Table below illustrates the levels of the indicator DISC in the MMRM3 model in the PIONEER Simulation.

| Pattern (B = Baseline: O = No IE: X = IE) | Description | Pattern Codes | | | | |
|---|---|---|---|---|---|---|
| | | V1 | V2 | V3 | V4 | V5 |
| BXXXXX | IE after Visit 0 (Post IE at Visit 1) | 1 | 1 | 1 | 1 | 1 |
| BOXXXX | IE after Visit 1 (Post IE at Visit 2) | 6 | 2 | 2 | 2 | 2 |
| BOOXXX | IE after Visit 2 (Post IE at Visit 3) | 6 | 6 | 3 | 3 | 3 |
| BOOOXX | IE after Visit 3 (Post IE at Visit 4) | 6 | 6 | 6 | 4 | 4 |
| BOOOOX | IE after Visit 4 (Post IE at Visit 5) | 6 | 6 | 6 | 6 | 5 |
| BOOOOO | IE does not occur (Completers) | 6 | 6 | 6 | 6 | 6 |
| | Total Covariate Levels | 2 | 3 | 4 | 5 | 6 |

# Appendix 2: Variance Adjustment for Treatment Policy

The variance of the treatment policy has components for the proportion in each pattern and the outcomes in each pattern. Below is a sketch derivation of the total variance calculation obtained by the delta method.

Assuming we have $j = 1, \dots, J$ visits we let $\hat{\theta}_j^{(p)}$ be the estimated proportion and $\hat{\mu}_j^{(p)}$ be the estimated mean of patients with IE occurrence pattern $p$ at visit $j$ with $p = 1, \dots, j+1$.

The expected values of the treatment policy can be estimated as:

$$E\left[Y_j^{policy}\right] = \hat{\mu}_j^{policy} = \hat{\theta}_j^{(1)} \cdot \hat{\mu}_j^{(1)} + \dots + \hat{\theta}_j^{(j)} \cdot \hat{\mu}_j^{(j)} + \left(1 - \sum_{p=1}^{j} \hat{\theta}_j^{(p)}\right) \cdot \hat{\mu}_j^{(j+1)}$$

The variance of the treatment policy expected value can be calculated using the delta method.

$$Var\left[\hat{\mu}_j^{policy}\right] = \nabla\left(\hat{\mu}_j^{policy}\right)^T \cdot \hat{\Sigma} \cdot \nabla\left(\hat{\mu}_j^{policy}\right)$$

Where:

$$V\begin{bmatrix} \theta_j^{(1)} \\ \vdots \\ \theta_j^{(j)} \\ \mu_j^{(1)} \\ \vdots \\ \mu_j^{(j)} \\ \mu_j^{(j+1)} \end{bmatrix} \approx \hat{\Sigma} = \begin{bmatrix} V\left(\hat{\theta}_j^{(1)}\right) & \cdots & C\left(\hat{\theta}_j^{(1)},\hat{\theta}_j^{(j)}\right) & 0 & \cdots & 0 & 0 \\ \vdots & \ddots & \vdots & \vdots & \ddots & \vdots & \vdots \\ C\left(\hat{\theta}_j^{(j)},\hat{\theta}_j^{(1)}\right) & \cdots & V\left(\hat{\theta}_j^{(j)}\right) & 0 & \cdots & 0 & 0 \\ 0 & \cdots & 0 & V\left(\hat{\mu}_j^{(1)}\right) & \cdots & C\left(\hat{\mu}_j^{(1)},\hat{\mu}_j^{(j)}\right) & C\left(\hat{\mu}_j^{(1)},\hat{\mu}_j^{(j+1)}\right) \\ \vdots & \ddots & \vdots & \vdots & \ddots & \vdots & \vdots \\ 0 & \cdots & 0 & C\left(\hat{\mu}_j^{(j)},\hat{\mu}_j^{(1)}\right) & \cdots & V\left(\hat{\mu}_j^{(j)}\right) & C\left(\hat{\mu}_j^{(j)},\hat{\mu}_j^{(j+1)}\right) \\ 0 & \cdots & 0 & C\left(\hat{\mu}_j^{(j+1)},\hat{\mu}_j^{(1)}\right) & \cdots & C\left(\hat{\mu}_j^{(j+1)},\hat{\mu}_j^{(j)}\right) & V\left(\hat{\mu}_j^{(j+1)}\right) \end{bmatrix}$$

With $V(\cdot)$ and $C(\cdot,\cdot)$ denoting the Variance and Covariance respectively. This makes the simplifying assumption that the variance and covariances of the proportions in each pattern are independent of the variance and covariances of the expected values in each pattern.

Also:

$$\nabla\left(\hat{\mu}_j^{policy}\right) = \left[\frac{\partial}{\partial \hat{\theta}_j^{(1)}}\hat{\mu}_j^{policy} \quad \cdots \quad \frac{\partial}{\partial \hat{\theta}_j^{(j)}}\hat{\mu}_j^{policy} \quad \frac{\partial}{\partial \hat{\mu}_j^{(1)}}\hat{\mu}_j^{policy} \quad \cdots \quad \frac{\partial}{\partial \hat{\mu}_j^{(j)}}\hat{\mu}_j^{policy} \quad \frac{\partial}{\partial \hat{\mu}_j^{(j+1)}}\hat{\mu}_j^{policy}\right]^T$$

Which is:

$$\nabla\left(\hat{\mu}_j^{policy}\right) = \left[\left(\hat{\mu}_j^{(1)} - \hat{\mu}_j^{(j+1)}\right) \quad \cdots \quad \left(\hat{\mu}_j^{(j)} - \hat{\mu}_j^{(j+1)}\right) \quad \hat{\theta}_j^{(1)} \quad \cdots \quad \hat{\theta}_j^{(j)} \quad \left(1 - \sum_{p=1}^{j} \hat{\theta}_j^{(p)}\right)\right]^T$$

Because of the independence assumption $\nabla\left(\hat{\mu}_j^{policy}\right)^T \cdot \hat{\Sigma} \cdot \nabla\left(\hat{\mu}_j^{policy}\right)$ can be broken into two distinct parts, the first for the variability in the pattern proportions, the second for the variability in the expected values for each pattern.

Specifically, the component for the variability of the proportions is:

$$V\begin{bmatrix}\theta_j^{(1)}\\\vdots\\\theta_j^{(j)}\end{bmatrix}=\begin{bmatrix}(\hat{\mu}_j^{(1)}-\hat{\mu}_j^{(j+1)}) & \cdots & (\hat{\mu}_j^{(j)}-\hat{\mu}_j^{(j+1)})\end{bmatrix}\begin{bmatrix}V(\hat{\theta}_j^{(1)}) & \cdots & C(\hat{\theta}_j^{(1)},\hat{\theta}_j^{(j)})\\\vdots & \ddots & \vdots\\C(\hat{\theta}_j^{(j)},\hat{\theta}_j^{(1)}) & \cdots & V(\hat{\theta}_j^{(j)})\end{bmatrix}\begin{bmatrix}(\hat{\mu}_j^{(1)}-\hat{\mu}_j^{(j+1)})\\\vdots\\(\hat{\mu}_j^{(j)}-\hat{\mu}_j^{(j+1)})\end{bmatrix}$$

$$V\begin{bmatrix}\theta_j^{(1)}\\\vdots\\\theta_j^{(j)}\end{bmatrix}=\begin{bmatrix}(\hat{\mu}_j^{(1)}-\hat{\mu}_j^{(j+1)}) & \cdots & (\hat{\mu}_j^{(j)}-\hat{\mu}_j^{(j+1)})\end{bmatrix}\begin{bmatrix}\dfrac{\hat{\theta}_j^{(1)}(1-\hat{\theta}_j^{(1)})}{n_j} & \cdots & -\dfrac{\hat{\theta}_j^{(1)}\hat{\theta}_j^{(j)}}{n_j}\\\vdots & \ddots & \vdots\\-\dfrac{\hat{\theta}_j^{(j)}\hat{\theta}_j^{(1)}}{n_j} & \cdots & \dfrac{\hat{\theta}_j^{(j)}(1-\hat{\theta}_j^{(j)})}{n_j}\end{bmatrix}\begin{bmatrix}(\hat{\mu}_j^{(1)}-\hat{\mu}_j^{(j+1)})\\\vdots\\(\hat{\mu}_j^{(j)}-\hat{\mu}_j^{(j+1)})\end{bmatrix}$$

And the component for the variability of the expected values is:

$$V\begin{bmatrix}\mu_j^{(1)}\\\vdots\\\mu_j^{(j)}\\\mu_j^{(j+1)}\end{bmatrix}=\begin{bmatrix}\hat{\theta}_j^{(1)} & \cdots & \hat{\theta}_j^{(j)} & \left(1-\sum_{p=1}^{j}\hat{\theta}_j^{(p)}\right)\end{bmatrix}\begin{bmatrix}V(\hat{\mu}_j^{(1)}) & \cdots & C(\hat{\mu}_j^{(1)},\hat{\mu}_j^{(j)}) & C(\hat{\mu}_j^{(1)},\hat{\mu}_j^{(j+1)})\\\vdots & \ddots & \vdots & \vdots\\C(\hat{\mu}_j^{(j)},\hat{\mu}_j^{(1)}) & \cdots & V(\hat{\mu}_j^{(j)}) & C(\hat{\mu}_j^{(j)},\hat{\mu}_j^{(j+1)})\\C(\hat{\mu}_j^{(j+1)},\hat{\mu}_j^{(1)}) & \cdots & C(\hat{\mu}_j^{(j+1)},\hat{\mu}_j^{(j)}) & V(\hat{\mu}_j^{(j+1)})\end{bmatrix}\begin{bmatrix}\hat{\theta}_j^{(1)}\\\vdots\\\hat{\theta}_j^{(j)}\\\left(1-\sum_{p=1}^{j}\hat{\theta}_j^{(p)}\right)\end{bmatrix}$$

The estimates for $V(\hat{\mu}_j^{(p)})$ and $C(\hat{\mu}_j^{(p)},\hat{\mu}_j^{(p')})$ can be extracted directly from the covariance matrix of the estimated parameters from the MMRM models. The total variance for the policy is then the sum of both components of the variance.

As the $k$ groups of patients are assumed independent the total variance for the estimated treatment policy effects between any of the $k$ groups at visit $j$ are:

$$Var\left[\hat{\mu}_{jk}^{policy}-\hat{\mu}_{jk'}^{policy}\right]=Var\left[\hat{\mu}_{jk}^{policy}\right]+Var\left[\hat{\mu}_{jk'}^{policy}\right]\quad :k\neq k'$$

# Appendix 3: RD Fitting Algorithm

The pattern fitting issues for the RD approaches result from patients discontinuing assigned treatment and then withdrawing from study at the current or future visits. Specifically, the problems are:

1. If no patients discontinue at a particular visit there is no need for a pattern level for that visit or future visits. This is not a large problem as it just means we need fewer overall discontinuation patterns.

2. If at least one patient discontinues at a visit then a pattern level is needed. But if all patients from this pattern withdraw from study at this visit (or a future visit) there will be no patients remaining to estimate the effects for that pattern. To address this, we combine these discontinuation patterns with other adjacent patterns (prioritising the previous pattern over the future pattern) to allow estimation.

3. If by a certain visit, there are no patients left in the trial for a particular discontinuation pattern, there may also be no other discontinuation patterns at that visit remaining (or occurred yet) to combine patterns, leaving only the remaining "on-treatment" patients. An example of this would be all patients that discontinue at visit 1 also withdraw from the study. Because no other discontinuation patterns exist yet at this visit, there are only "on-treatment" patients. In this situation, further borrowing of discontinuation outcomes could occur from future discontinuation patterns, but given the increased complexity of trying to borrow information from future patterns, we opt to use the remaining "on-treatment" patients at the same visit.

To ensure all the RD models can be estimated, we need to check these issues to work out if patterns need to be combined to allow estimation. For a single analysis, this could be done by manually checking the data, however for our simulation study, this needs to be automated as an algorithm. To do this we create three sets of indicators labelled: pattern issue, data issue and estimation issue described below:

- The pattern issue indicators ($p\_i1$-$p\_i5$) check **each discontinuation pattern** (1 to 5). Each pattern indicator is set to one if there is at least one patient providing data at all the required visits for the pattern (e.g. patients in discontinuation pattern 1 need at least one patient to provide data at all 5 visits, whereas at the other extreme, patients with discontinuation pattern 5 requires one patient to provide data at the final visit only).
- The data issue indicators ($d\_i1$-$d\_i5$) check there is at least one discontinuation pattern providing data at **each visit** (1 to 5). Each indicator is set to one if there are some discontinuation data at the visit which can therefore be used when combining patterns.
- The estimation issue indicators ($e\_i1$-$e\_i5$) combine the pattern issue and data issue indicators. They are set to one for any situations where there is no pattern data for a discontinuation pattern, but also no other discontinuation pattern data at the visit and therefore no discontinuation pattern terms can be estimated at that visit.

To assess issue #2, the pattern issue indicators (p_i1-p_i5) are arranged into a set of binary indicators and then the simplification algorithim applied to the cases with pattern issues. The combining of patterns results in a recoding of the discontinuation status used in the MMRM2/MI2 models and discontinuation pattern used in the MMRM3/MI3 models. The overall approach is summarised in Table X below:

*Table A5: Summary of the pattern combining that occurs in the simplification algorithim to ensure the RD models can be fitted. The recoded discontinuation codes are used for the MMRM2 and MI2 models and the recoded pattern codes are used for the MMRM3 and MI3 models.*

| Pattern Issue Indicators | | | | | Max. Patterns | Recoded Disc. Code | Recoded Pat. Code |
|---|---|---|---|---|---|---|---|
| p_i1 | p_i2 | p_i3 | p_i4 | p_i5 | | | |
| 0 | 0 | 0 | 0 | 0 | 6 Patterns | 12345, 6 | 1, 2, 3, 4, 5, 6 |
| 1 | 0 | 0 | 0 | 0 | 5 Patterns | 12345, 6 | 12, 3, 4, 5, 6 |
| 0 | 1 | 0 | 0 | 0 | | 12345, 6 | 12, 3, 4, 5, 6 |
| 0 | 0 | 1 | 0 | 0 | | 12345, 6 | 1, 23, 4, 5, 6 |
| 0 | 0 | 0 | 1 | 0 | | 12345, 6 | 1, 2, 34, 5, 6 |
| 0 | 0 | 0 | 0 | 1 | | 12345, 6 | 1, 2, 3, 45, 6 |
| 1 | 1 | 0 | 0 | 0 | 4 Patterns | 12345, 6 | 123, 4, 5, 6 |
| 1 | 0 | 1 | 0 | 0 | | 12345, 6 | 123, 4, 5, 6 |
| 0 | 1 | 1 | 0 | 0 | | 12345, 6 | 123, 4, 5, 6 |
| 0 | 0 | 1 | 1 | 0 | | 12345, 6 | 1, 234, 5, 6 |
| 0 | 0 | 0 | 1 | 1 | | 12345, 6 | 1, 2, 345, 6 |
| 1 | 0 | 0 | 1 | 0 | | 12345, 6 | 12, 34, 5, 6 |
| 0 | 1 | 0 | 1 | 0 | | 12345, 6 | 12, 34, 5, 6 |
| 1 | 0 | 0 | 0 | 1 | | 12345, 6 | 12, 3, 45, 6 |
| 0 | 1 | 0 | 0 | 1 | | 12345, 6 | 12, 3, 45, 6 |
| 0 | 0 | 1 | 0 | 1 | | 12345, 6 | 1, 23, 45, 6 |
| 1 | 1 | 0 | 1 | 0 | 3 Patterns | 12345, 6 | 1234, 5, 6 |
| 1 | 0 | 1 | 1 | 0 | | 12345, 6 | 1234, 5, 6 |
| 1 | 1 | 1 | 0 | 0 | | 12345, 6 | 1234, 5, 6 |
| 0 | 1 | 1 | 1 | 0 | | 12345, 6 | 1234, 5, 6 |
| 0 | 0 | 1 | 1 | 1 | | 12345, 6 | 1, 2345, 6 |
| 1 | 1 | 0 | 0 | 1 | | 12345, 6 | 123, 45, 6 |
| 1 | 0 | 1 | 0 | 1 | | 12345, 6 | 123, 45, 6 |
| 0 | 1 | 1 | 0 | 1 | | 12345, 6 | 123, 45, 6 |
| 1 | 0 | 0 | 1 | 1 | | 12345, 6 | 12, 345, 6 |
| 0 | 1 | 0 | 1 | 1 | | 12345, 6 | 12, 345, 6 |
| 1 | 1 | 1 | 1 | 0 | 2 Patterns | 12345, 6 | 12345, 6 |
| 1 | 1 | 1 | 0 | 1 | | 12345, 6 | 12345, 6 |
| 1 | 1 | 0 | 1 | 1 | | 12345, 6 | 12345, 6 |
| 1 | 0 | 1 | 1 | 1 | | 12345, 6 | 12345, 6 |
| 0 | 1 | 1 | 1 | 1 | | 12345, 6 | 12345, 6 |
| 1 | 1 | 1 | 1 | 1 | 1 Pattern | 123456 | 123456 |

Although theoretically these indicate which patterns should be combined, for any models with more than one pattern, issue #3 relating to the need to combine a pattern at a visit when there is no other discontinuation data available also needs checking. This checked using a combination of the pattern indicators and estimation indicators. For situations that permit 2 or more patterns, any visit where the there is a need to combine discontinuation patterns but there are no data from other discontinuation patterns (remaining or occurred yet), we set the discontinuation or pattern code to "on-treatment" (level 6).

# Appendix 4: Simulation setup

## Relevant PIONEER 1 results

Table A2 presents the on-study means, sample sizes, and standard deviations for the baseline visit (week 0; W0) and the post-baseline visits (W4-W26) for HbA1c in the PIONEER 1 study. Analogously, Table A3 presents the corresponding on-treatment means, sample sizes, and standard deviations for HbA1c.

*Table A6 On-study means, sample sizes, and standard deviations for HbA1c. This includes patients regardless of discontinuation of randomized trial product and use of rescue medication.*

| RANDOMIZED TREATMENT | PARAMETER | W0 | W4 | W8 | W14 | W20 | W26 |
|---|---|---|---|---|---|---|---|
| PLACEBO | Mean | 7.9146 | 7.7937 | 7.7524 | 7.7906 | 7.7151 | 7.6190 |
|  | N | 178 | 174 | 170 | 171 | 166 | 168 |
|  | SD | 0.6792 | 0.8895 | 1.0780 | 1.1965 | 1.1713 | 1.2410 |
| SEMAGLUTIDE 3 MG | Mean | 7.9240 | 7.5506 | 7.1793 | 7.0876 | 7.0200 | 7.0347 |
|  | N | 175 | 172 | 169 | 169 | 165 | 167 |
|  | SD | 0.7045 | 0.8654 | 0.9325 | 0.9924 | 1.0409 | 1.0639 |

*Table A7 "On-trial-product" means, sample sizes, and standard deviations for HbA1c. Only outcomes prior to discontinuation of randomized trial product and use of rescue medication are included.*

| RANDOMIZED TREATMENT | PARAMETER | W0 | W4 | W8 | W14 | W20 | W26 |
|---|---|---|---|---|---|---|---|
| PLACEBO | Mean | 7.9146 | 7.8023 | 7.7382 | 7.7411 | 7.5993 | 7.5158 |
|  | N | 178 | 172 | 165 | 158 | 140 | 133 |
|  | SD | 0.6792 | 0.8909 | 1.0492 | 1.1797 | 1.1098 | 1.2182 |
| SEMAGLUTIDE 3 MG | Mean | 7.9240 | 7.5580 | 7.1390 | 7.0032 | 6.9364 | 6.9779 |
|  | N | 175 | 169 | 164 | 156 | 151 | 149 |
|  | SD | 0.7045 | 0.8612 | 0.8891 | 0.9467 | 1.0311 | 1.0665 |

## Simulating patient-level data

Our simulation of patient-level clinical trial data comprises four steps: The first step is to simulate data for 200 patients per arm under the hypothetical setting that all patients remain on their initially assigned trial product. Secondly, the occurrences of any intercurrent events are simulated. Next, for those patients who experience the intercurrent event, the subsequent off-trial-product response is simulated. Lastly, the missingness indicators are generated and the missing values removed. Details for the respective steps are provided in the following.

The focus is on a two-arm randomized clinical trial with an experimental treatment arm $T$ and a control arm $C$. As for the notation, we use $k = T, C$ as the index for the group, $i = 1, \ldots, n_k$ as the index for the patient within treatment arm, and $j = 0, \ldots, J$ as the index for the visit time point with $j = 0$ for the baseline visit and $j > 0$ for post-baseline visits.

**Simulation model for data in absence of intercurrent events**

Let $\boldsymbol{Y}_{ki} = (Y_{ki0}, \ldots, Y_{kiJ})'$ be the random vector modelling the outcome of patient $i$ in group $k$ across visits for the case that the patient continues to take the initially assigned treatment. The vector $\boldsymbol{Y}_{ki}$ follows a multivariate normal distribution

$$\boldsymbol{Y}_{ki} \sim MVN(\boldsymbol{\mu}_k, \boldsymbol{\Sigma}_k)$$

with $\boldsymbol{\mu}_k = (\mu_{k0}, \ldots, \mu_{kJ})$ the vector of means across visits, and $\boldsymbol{\Sigma}_k$ the covariance matrix. The values for the parameters are provided in Table A4.

*Table A8 Parameter choices for the on-trial-product means and variances. The variance matrices $\boldsymbol{\Sigma}_T$ and $\boldsymbol{\Sigma}_C$ and created using a first-order spatial power structure and a correlation of $\rho^{|t_i-t_j|/|t_0-t_1|}$ with $t_i$ the visit in weeks and $\rho = 0.8$.*

| Visit (week) | Treatment group mean $\mu_T$ | Control group mean $\mu_C$ | Treatment group variances $diag(\Sigma_T)$ | Control group variances $diag(\Sigma_C)$ |
|---|---|---|---|---|
| 0 | 7.92 | 7.92 | 0.48 | 0.48 |
| 4 | 7.55 | 7.82 | 0.75 | 0.8 |
| 8 | 7.2 | 7.8 | 0.8 | 1.1 |
| 14 | 7.1 | 7.8 | 0.9 | 1.4 |
| 20 | 7.05 | 7.78 | 1.06 | 1.23 |
| 26 | 7.05 | 7.78 | 1.14 | 1.48 |

**Simulation model for trial product discontinuation**

Let $\boldsymbol{D}_{ki} = (D_{ki1}, \ldots, D_{kiJ})$ be the vector of discontinuation indicators for the post-baseline visits. $D_{kij}$ is equal to 1 if patient $i$ in group $k$ has discontinued their randomized trial product up to visit $j$, that is $D_{kij_1} = 1$ for all $j_1 \geq j$ with $D_{kij} = 1$. Otherwise, $D_{kij} = 0$. The outcomes at visits with $D_{kij} = 1$ are those that are affected by the discontinuation; see below for details.

The distribution of the treatment discontinuation profile is specified recursively through conditional probabilities

$$p_{kij}^D = P\left(D_{kij} = 1 \middle| D_{ki(j-1)} = 0, (Y_{ki0}, \ldots, Y_{kij})\right).$$

We consider two models for treatment discontinuation, which we name following the missing data framework language [20]. In the first model, which we denote as the 'discontinuation at random' (DAR) model, the probability $p_{kij}^D$ depends on the group $k$, the visit $j$, the baseline observation $Y_{ki0}$, and the observation $Y_{ki(j-1)}$ of the preceding visit. In the second model, which we denote as the 'discontinuation not at random' (DNAR) model, the probability $p_{kij}^D$ depends on the same parameters as in the DAR model and additionally also depends on the observation $Y_{kij}$. The motivation for modelling the probability of discontinuation dependent on $Y_{kij}$ is to capture the effect of unmeasured declining health/response to the trial product in the time since the previous visit.

*DAR model for trial product discontinuation*

In detail, in the DAR model for treatment discontinuation, the probability $p_{kij}^D$ is affected by the different factors through a logit-link function, that is

$$\text{logit}(p_{kij}^D) = \beta_{kj}^{(0)} + \beta_{kj}^{(1)} Y_{ki0} + \beta_{kj}^{(2)} Y_{ki(j-1)}.$$

*DNAR model for trial product discontinuation*

In comparison to the DAR model, the DNAR include one additional term that defines the dependence of the probability $p_{kij}^D$ on the (potential) on-treatment value $Y_{kij}$:

$$\text{logit}(p_{kij}^D) = \beta_{kj}^{(0)} + \beta_{kj}^{(1)} Y_{ki0} + \beta_{kj}^{(2)} Y_{ki(j-1)} + \beta_{kj}^{(3)} Y_{kij}.$$

The parameter choices are provided in Table A5 and Table A6.

*Table A9 Parameter choices for the DAR model for trial product discontinuation.*

| Parameter | Treatment arm ($k = T$) | Control arm ($k = C$) |
|---|---|---|
| $\beta_{kj}^{(0)}; j = 1, \ldots, 5$ | -15 | -15 |
| $\beta_{k1}^{(1)}$ | 0 | 0 |
| $\beta_{k2}^{(1)}$ | 0.3 | 0.3 |
| $\beta_{k3}^{(1)}$ | 0.1 | 0.1 |
| $\beta_{k4}^{(1)}$ | 0.05 | 0.05 |
| $\beta_{k5}^{(1)}$ | 0 | 0 |
| $\beta_{k1}^{(2)}$ | 1.42 | 1.42 |
| $\beta_{k2}^{(2)}$ | 1.14 | 1.14 |
| $\beta_{k3}^{(2)}$ | 1.47 | 1.33 |
| $\beta_{k4}^{(2)}$ | 1.48 | 1.51 |
| $\beta_{k5}^{(2)}$ | 1.40 | 1.46 |

*Table A10 Parameter choices for the DNAR model for trial product discontinuation*

| Parameter | Treatment arm ($k = T$) | Control arm ($k = C$) |
|---|---|---|
| $\beta_{kj}^{(0)}; j = 1, \ldots, 5$ | -21 | -21 |
| $\beta_{k1}^{(1)}$ | 0 | 0 |
| $\beta_{k2}^{(1)}$ | 0.3 | 0.3 |
| $\beta_{k3}^{(1)}$ | 0.1 | 0.1 |
| $\beta_{k4}^{(1)}$ | 0.05 | 0.05 |
| $\beta_{k5}^{(1)}$ | 0 | 0 |
| $\beta_{k1}^{(2)}$ | 1.42 | 1.42 |
| $\beta_{k2}^{(2)}$ | 1.14 | 1.14 |
| $\beta_{k3}^{(2)}$ | 1.47 | 1.33 |
| $\beta_{k4}^{(2)}$ | 1.48 | 1.51 |
| $\beta_{k5}^{(2)}$ | 1.40 | 1.46 |
| $\beta_{k1}^{(3)}$ | 0.72 | 0.69 |
| $\beta_{k2}^{(3)}$ | 0.75 | 0.69 |
| $\beta_{k3}^{(3)}$ | 0.77 | 0.69 |
| $\beta_{k4}^{(3)}$ | 0.77 | 0.72 |
| $\beta_{k5}^{(3)}$ | 0.76 | 0.72 |

**Simulation model for off-trial-product data**

The random vector $Y_{ki}$ denotes a patient's trajectory under the assumption that the patient remains on their randomized trial product. We define the random vector $\widetilde{Y}_{ki}$ as that patient's trajectory under the possibility that the patient can discontinue their randomized treatment. For this purpose, let $\tau_{ki}$ be the time of the first visit affected by treatment discontinuation for patient $i$ in group $k$, that is $\tau_{ki} = \min\{j: D_{kij} = 1\}$. Moreover, let $s_{ki}(j)$ be the time since treatment discontinuation of patient $i$ in group $k$ at visit $j$, that is $s_{ki}(j) = j - \tau_{ki}$. Then, we define the entries of $\widetilde{Y}_{ki}$ through a linear shift of the entries of $Y_{ki}$ once the patient discontinues treatment:

$$\tilde{Y}_{kij} = \begin{cases} Y_{kij}, & j < \tau_{ki} \\ Y_{kij} + \delta_k(s_{ki}(j)), & j \geq \tau_{ki} \end{cases}$$

We distinguish two models for $\delta_k(\cdot)$. In the first model, denoted "Instant change", $\delta_k(\cdot)$ is a constant. A negative value corresponds to an immediate improvement of HbA1c at the time of the intercurrent event which is aligned with the assumption in our simulation study that the intercurrent event corresponds to patients simultaneously discontinuing the trial product and taking recue medication. In the second model, denoted "Gradual change", $\delta_k(\cdot)$ has the form $\delta_k(x) = a \cdot \min(x, b)/b$ which corresponds to a shift of HbA1c that increases over $b$ visits up to a shift of $a$. The definitions of $\delta_k(x)$ for the simulation study are provided in the appendix [add reference to table in appendix later].

*Table A11 Definition of function $\delta_k(\cdot)$ which defines the off-trial-product means.*

| Model | Treatment arm ($k = T$) | Control arm ($k = C$) |
|---|---|---|
| Immediate change | $\delta_T(x) = -0.2$ | $\delta_C(x) = -0.6$ |
| Gradual change | $\delta_T(x) = -0.25 \cdot \min(x, 3)/3$ | $\delta_C(x) = -0.8 \cdot \min(x, 3)/3$ |

**Simulation model for missingness**

As a last step, the model for missingness is defined. Let $\boldsymbol{C}_{ki} = (C_{ki1}, \ldots, C_{kiJ})$ be the vector of missingness indicators for patient $i$ in group $k$. $C_{kij} = 1$ indicates that $\tilde{Y}_{kij}$ is missing and analogously $C_{kij} = 0$ indicates that $\tilde{Y}_{kij}$ is observed. We assume that missingness is not intermittent, that is once a patient has a missing observation, the observations from all future visits will also be missing: $C_{k_1ij} = 1$ for all $k_1 > k$ with $C_{kij} = 1$. We also assume that missing observations can only occur for those patients who already discontinued the treatment.

The distribution of the missingness profile is specified recursively through conditional probabilities

$$p^C_{kij} = P(C_{kij} = 1 | C_{ki(j-1)} = 0, D_{kji}).$$

The probability $p^C_{kij}$ is defined as follows:

$$p^C_{kij} = \begin{cases} 0, & D_{kij} = 0 \\ expit(\theta_k^{(1)}), & D_{kij} = 1 \end{cases}$$

Here, $expit(x) = logit^{-1}(x) = \frac{1}{1+\exp(-x)}$ is the inverse of the logit-function. An overview of the numeric values of $\theta_k^{(1)}$ that are considered in the simulation is provided in Table A8.

*Table A12 Values of the parameters used in the missingness model.*

| Parameter | Values |
|---|---|
| $\theta_k^{(1)}$ | 0.1, 0.2, …, 0.6 |

## Comparison of the simulated data with PIONEER 1 data

Figure A1 shows the on-study means for the simulated data (solid line) and PIONEER 1 (dashed line). Figure A2 shows the on-treatment means for the simulated data (solid line) and PIONEER 1 (dashed line). For the simulated data, no missingness was simulated.

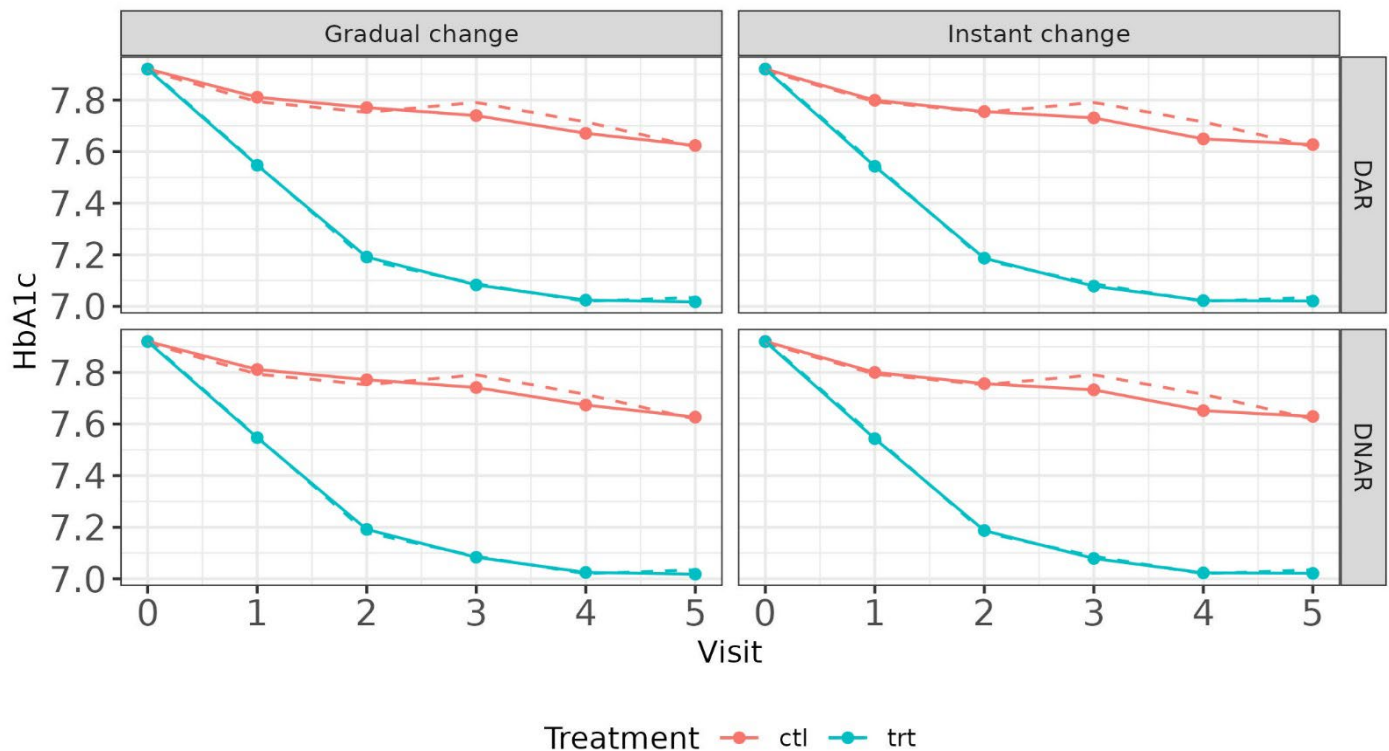

*Figure A6 On-study means for the simulated data (solid line) and PIONEER 1 (dashed line).*

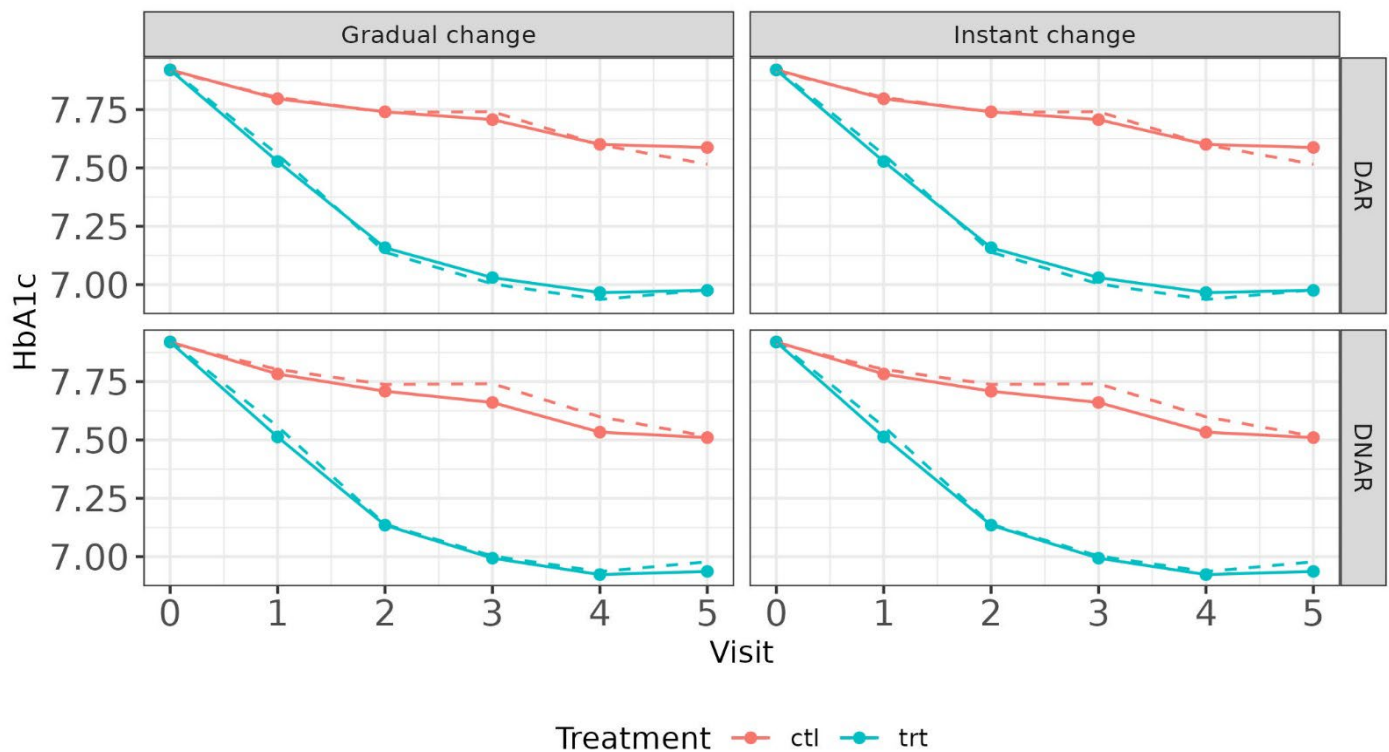

*Figure A7 On-treatment means for the simulated data (solid line) and PIONEER 1 (dashed line).*

**Supporting Information**

**Tables**

*Table SI1*: A breakdown of the missingness scenarios created, showing the average proportions of patients still on assigned treatment (O), the proportion that discontinued assigned treatment and still in the trial (D), or discontinued assigned treatment and subsequently withdrew from the trial (W).

| Missingness Scenario | Group | Status | Visit 1 | Visit 2 | Visit 3 | Visit 4 | Visit 5 |
|---|---|---|---|---|---|---|---|
| 1 | Control | O | 96.5 | 92.5 | 88.5 | 78.3 | 74.5 |
| 1 | | D | 3.1 | 6.4 | 9.4 | 17.6 | 19.3 |
| 1 | | W | 0.3 | 1.1 | 2.1 | 4.1 | 6.2 |
| 1 | Treatment | O | 96.5 | 93.6 | 88.9 | 86.0 | 84.8 |
| 1 | | D | 3.1 | 5.5 | 9.1 | 10.9 | 10.8 |
| 1 | | W | 0.3 | 1.0 | 2.0 | 3.2 | 4.4 |
| 2 | Control | O | 96.5 | 92.5 | 88.5 | 78.3 | 74.5 |
| 2 | | D | 2.8 | 5.4 | 7.5 | 14.2 | 14.4 |
| 2 | | W | 0.7 | 2.0 | 3.9 | 7.5 | 11.1 |
| 2 | Treatment | O | 96.5 | 93.6 | 88.9 | 86.0 | 84.8 |
| 2 | | D | 2.8 | 4.6 | 7.4 | 8.3 | 7.5 |
| 2 | | W | 0.7 | 1.8 | 3.7 | 5.8 | 7.7 |
| 3 | Control | O | 96.5 | 92.5 | 88.5 | 78.3 | 74.5 |
| 3 | | D | 2.4 | 4.5 | 6.0 | 11.3 | 10.6 |
| 3 | | W | 1.0 | 3.0 | 5.5 | 10.3 | 14.9 |
| 3 | Treatment | O | 96.5 | 93.6 | 88.9 | 86.0 | 84.8 |
| 3 | | D | 2.4 | 3.8 | 5.9 | 6.2 | 5.1 |
| 3 | | W | 1.0 | 2.6 | 5.2 | 7.9 | 10.1 |
| 4 | Control | O | 96.5 | 92.5 | 88.5 | 78.3 | 74.5 |
| 4 | | D | 2.1 | 3.7 | 4.6 | 8.9 | 7.6 |
| 4 | | W | 1.4 | 3.8 | 6.9 | 12.8 | 17.9 |
| 4 | Treatment | O | 96.5 | 93.6 | 88.9 | 86.0 | 84.8 |
| 4 | | D | 2.1 | 3.0 | 4.6 | 4.5 | 3.4 |
| 4 | | W | 1.4 | 3.4 | 6.5 | 9.5 | 11.8 |
| 5 | Control | O | 96.5 | 92.5 | 88.5 | 78.3 | 74.5 |
| 5 | | D | 1.7 | 2.9 | 3.4 | 6.8 | 5.3 |
| 5 | | W | 1.7 | 4.6 | 8.0 | 14.9 | 20.2 |
| 5 | Treatment | O | 96.5 | 93.6 | 88.9 | 86.0 | 84.8 |
| 5 | | D | 1.8 | 2.3 | 3.5 | 3.2 | 2.2 |
| 5 | | W | 1.7 | 4.1 | 7.6 | 10.8 | 13.0 |
| 6 | Control | O | 96.5 | 92.5 | 88.5 | 78.3 | 74.5 |
| 6 | | D | 1.4 | 2.2 | 2.5 | 5.0 | 3.5 |
| 6 | | W | 2.1 | 5.3 | 9.0 | 16.6 | 22.0 |
| 6 | Treatment | O | 96.5 | 93.6 | 88.9 | 86.0 | 84.8 |
| 6 | | D | 1.4 | 1.7 | 2.6 | 2.2 | 1.3 |
| 6 | | W | 2.1 | 4.7 | 8.5 | 11.8 | 13.9 |

*Table SI2: Proportion of discontinued patients that continued to be observed (D) or subsequently withdrew from the trial (W).*

| Missingness Scenario | Group | Status | Visit 1 | Visit 2 | Visit 3 | Visit 4 | Visit 5 |
|---|---|---|---|---|---|---|---|
| 1 | Control | D | 90.1 | 85.7 | 81.5 | 81.2 | 75.7 |
| 1 | | W | 9.9 | 14.3 | 18.5 | 18.8 | 24.3 |
| 1 | Treatment | D | 90.2 | 85.1 | 82.2 | 77.5 | 71.1 |
| 1 | | W | 9.8 | 14.9 | 17.8 | 22.5 | 28.9 |
| 2 | Control | D | 80.2 | 72.7 | 65.7 | 65.4 | 56.6 |
| 2 | | W | 19.8 | 27.3 | 34.3 | 34.6 | 43.4 |
| 2 | Treatment | D | 80.4 | 71.5 | 66.7 | 58.9 | 49.5 |
| 2 | | W | 19.6 | 28.5 | 33.3 | 41.1 | 50.5 |
| 3 | Control | D | 70.3 | 60.6 | 51.9 | 52.3 | 41.6 |
| 3 | | W | 29.7 | 39.4 | 48.1 | 47.7 | 58.4 |
| 3 | Treatment | D | 70.0 | 58.9 | 53.0 | 44.0 | 33.7 |
| 3 | | W | 30.0 | 41.1 | 47.0 | 56.0 | 66.3 |
| 4 | Control | D | 60.1 | 48.9 | 39.9 | 40.9 | 29.9 |
| 4 | | W | 39.9 | 51.1 | 60.1 | 59.1 | 70.1 |
| 4 | Treatment | D | 59.6 | 46.8 | 41.4 | 32.1 | 22.2 |
| 4 | | W | 40.4 | 53.2 | 58.6 | 67.9 | 77.8 |
| 5 | Control | D | 50.1 | 38.5 | 30.0 | 31.4 | 20.9 |
| 5 | | W | 49.9 | 61.5 | 70.0 | 68.6 | 79.1 |
| 5 | Treatment | D | 50.1 | 36.5 | 31.4 | 23.0 | 14.3 |
| 5 | | W | 49.9 | 63.5 | 68.6 | 77.0 | 85.7 |
| 6 | Control | D | 39.9 | 29.0 | 21.6 | 23.3 | 13.9 |
| 6 | | W | 60.1 | 71.0 | 78.4 | 76.7 | 86.1 |
| 6 | Treatment | D | 39.6 | 26.9 | 23.1 | 15.6 | 8.7 |
| 6 | | W | 60.4 | 73.1 | 76.9 | 84.4 | 91.3 |

*Figure SI1*: Treatment Effect Bias for DNAR mechanism (5000 simulations)

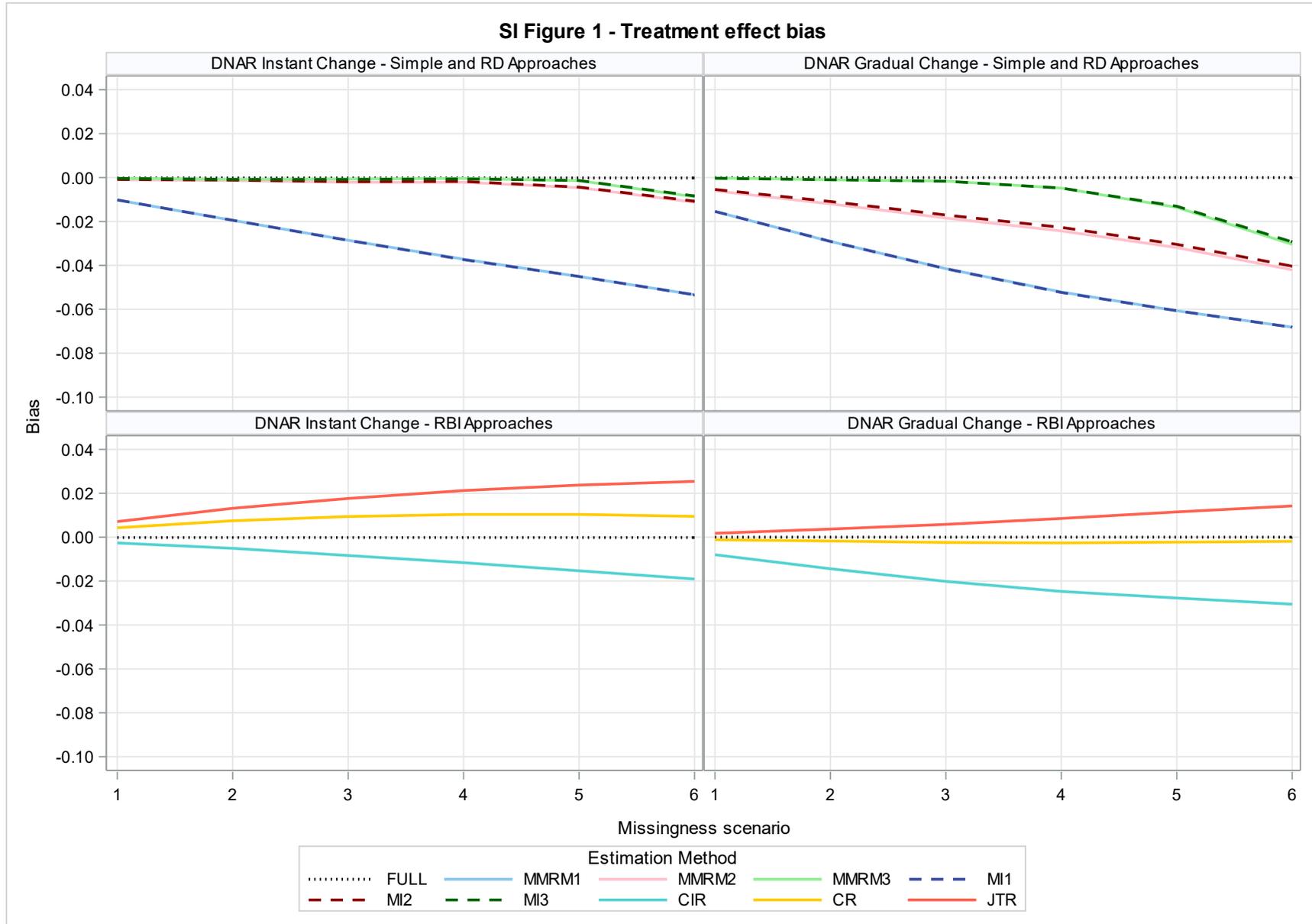

*Figure SI2*: SD of the treatment effect estimates for DNAR mechanism (5000 simulations)

*Figure SI3*: Model SE for treatment effects for DNAR mechanism. (5000 simulations)

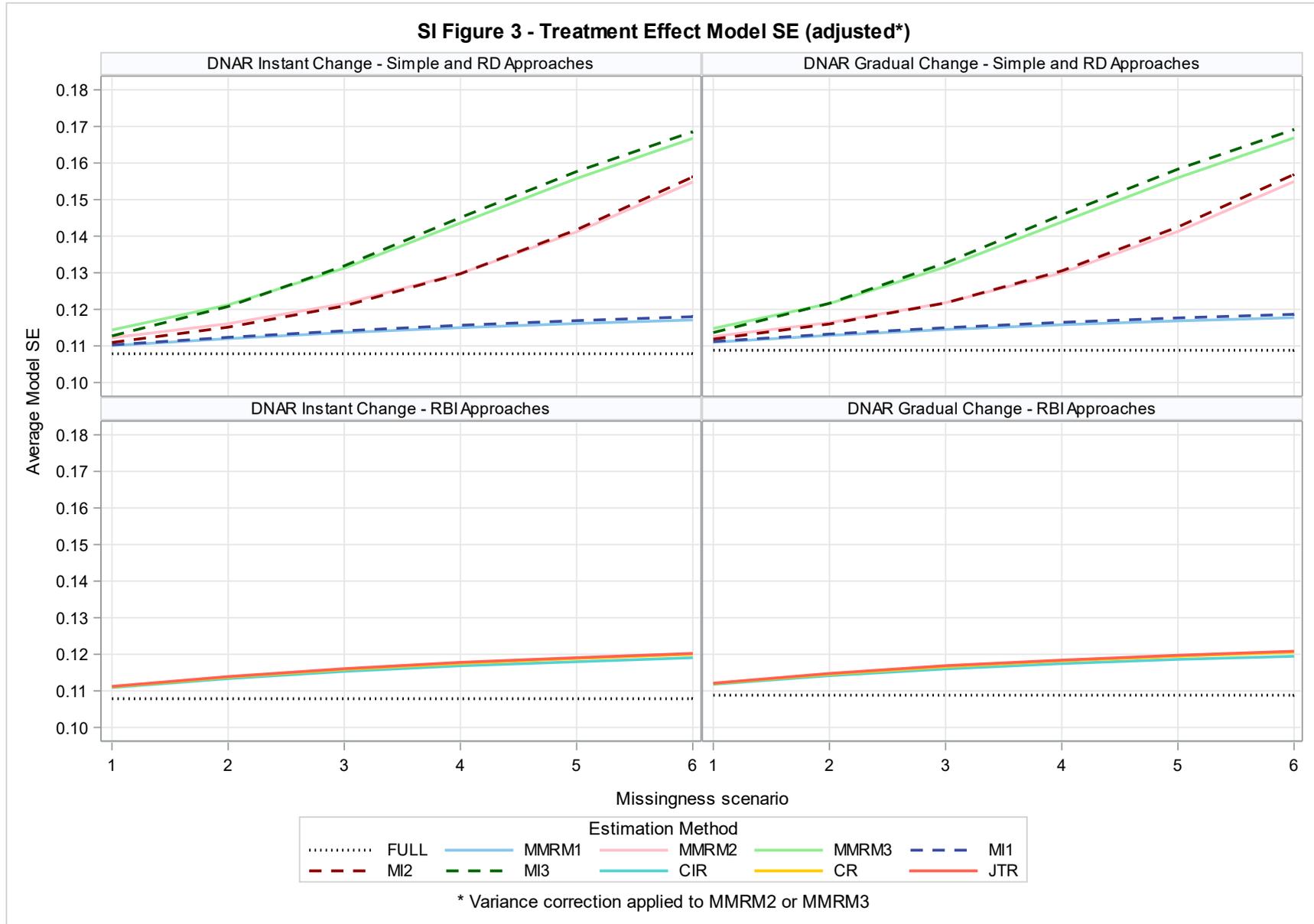

*Figure SI4*: Coverage of treatment effect 95% confidence intervals for DNAR mechanism. (5000 simulations)

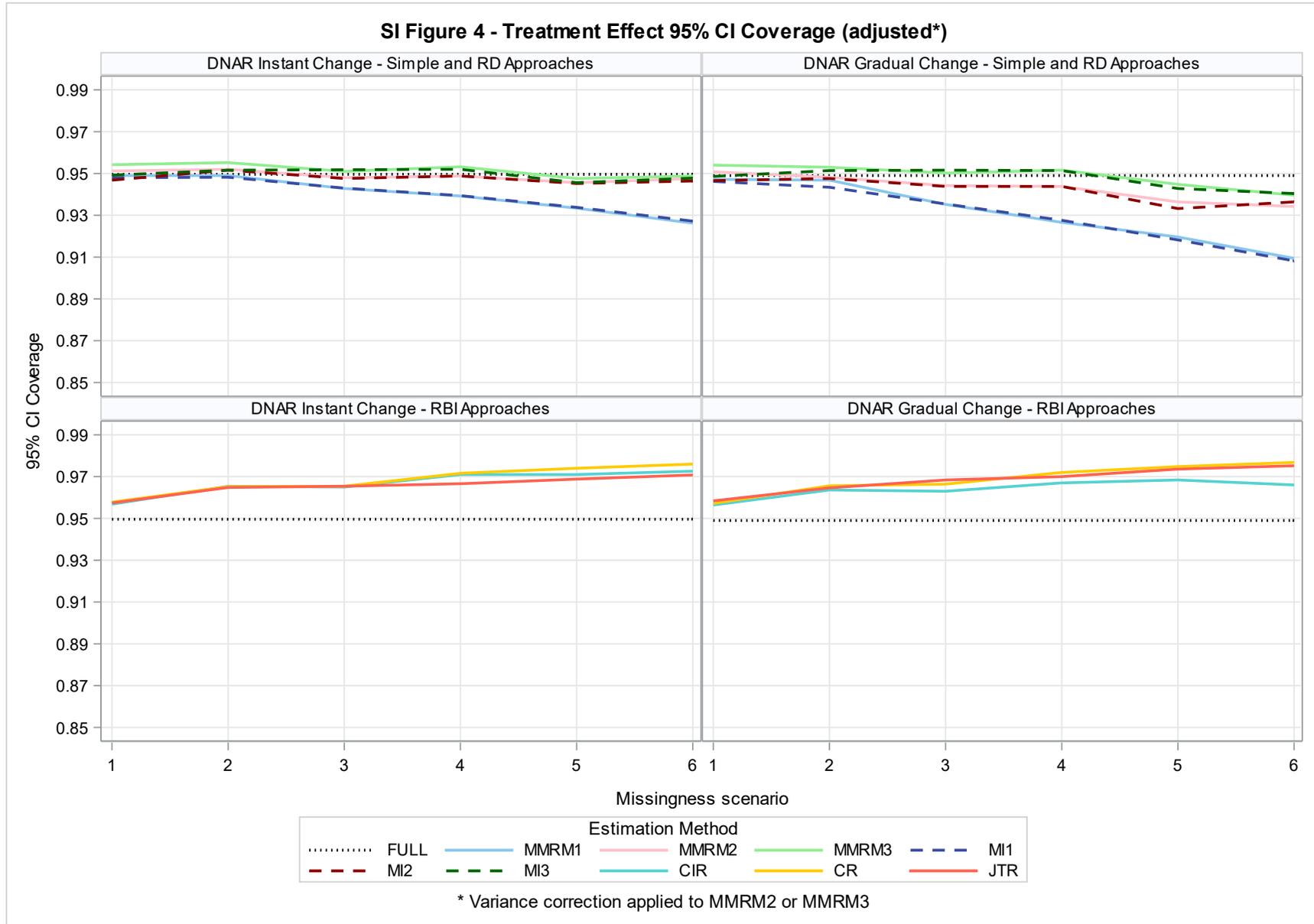

*Figure SI5*: Treatment effect root of average mean squared error for DAR mechanism. (5000 simulations)

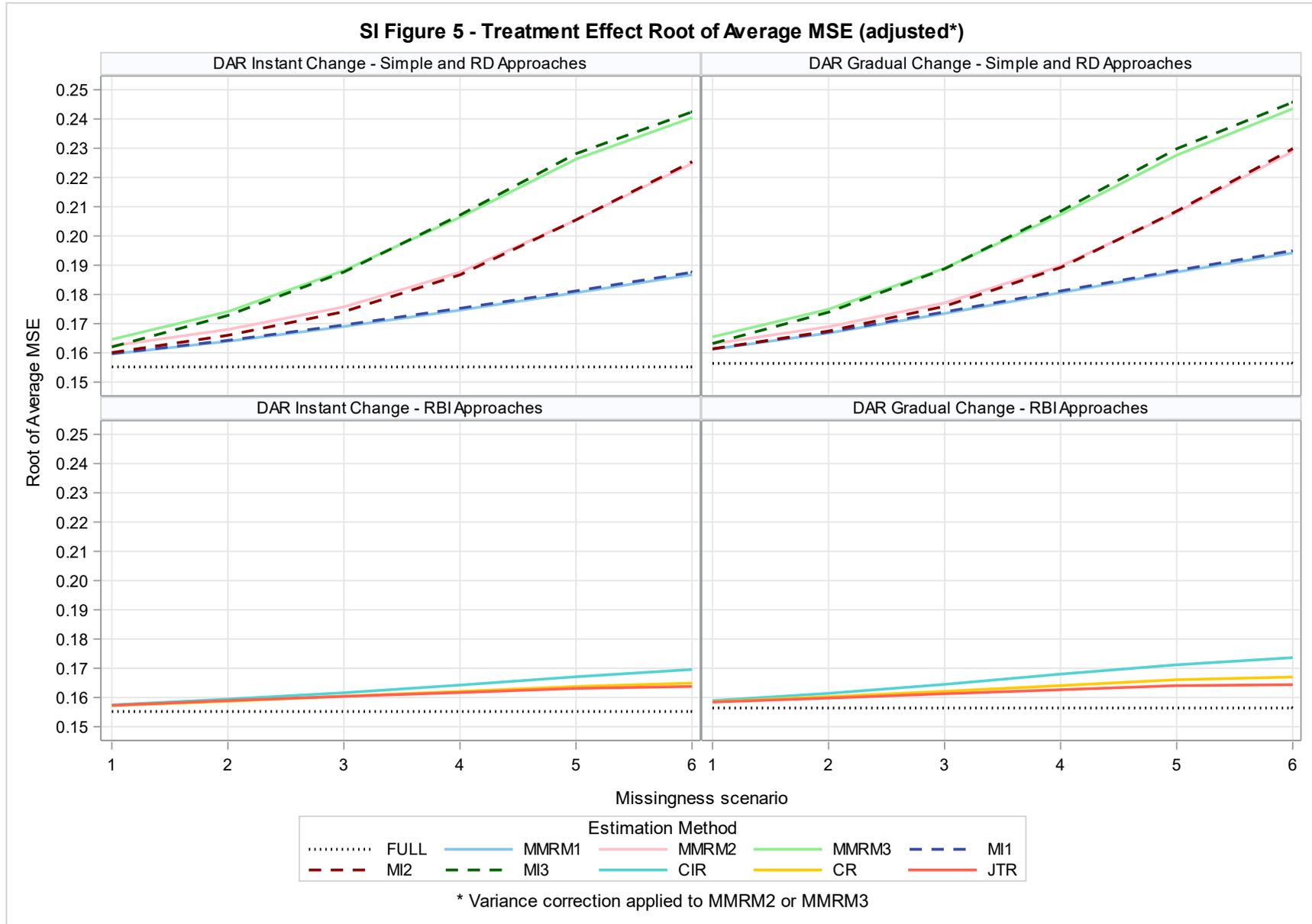

*Figure SI6*: Treatment effect root of average mean squared error for DNAR mechanism. (5000 simulations)

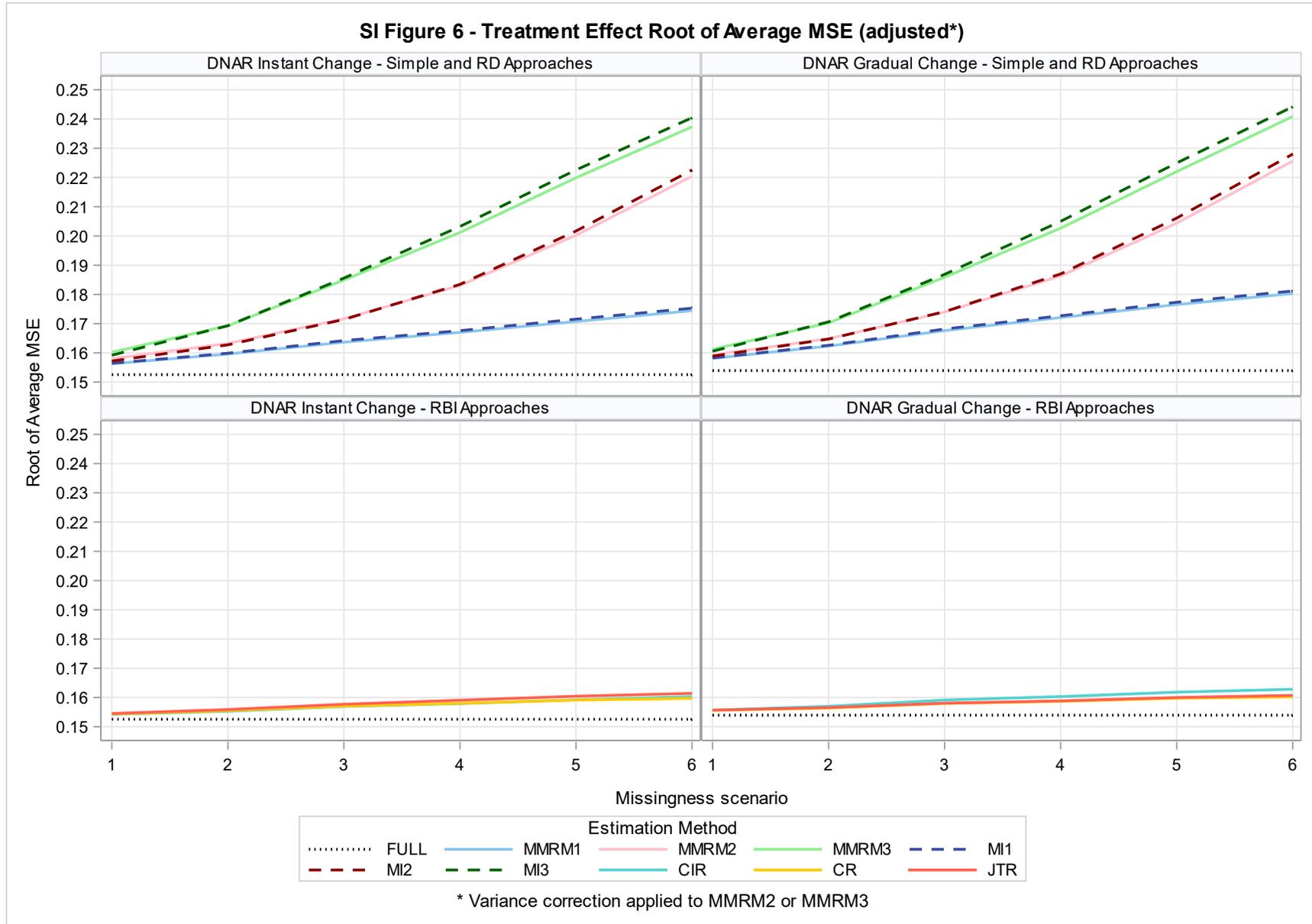

***Figure SI7.*** *Zipper plot for the DGM with discontinuation at random and gradual loss of treatment effect. The zipper plot is zoomed so that only the 15% of runs with theta further away from the true theta are shown. Confidence interval for coverage is shown is green (if compatible with 95% coverage) or red (if there is under- or over-coverage)*

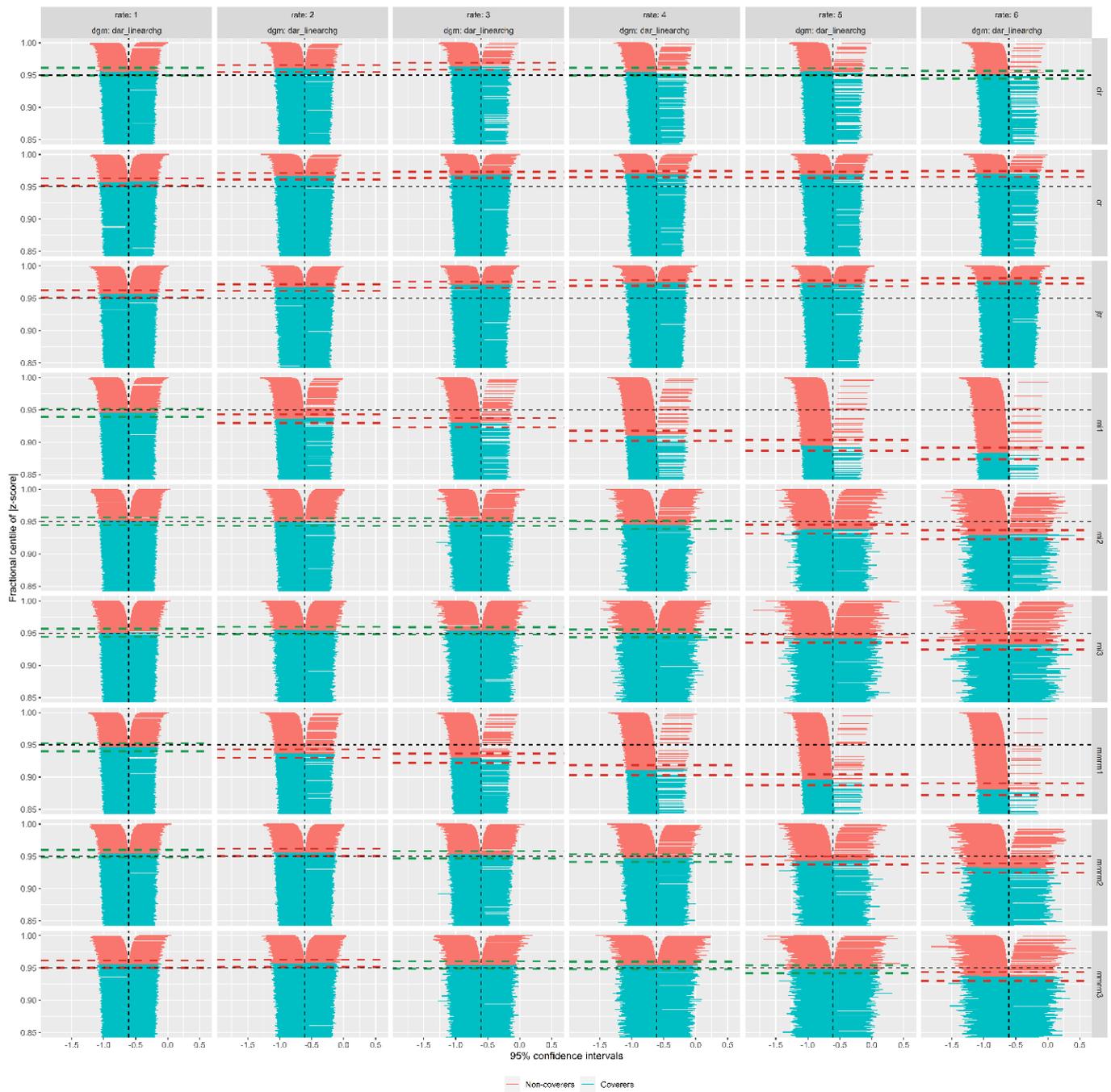

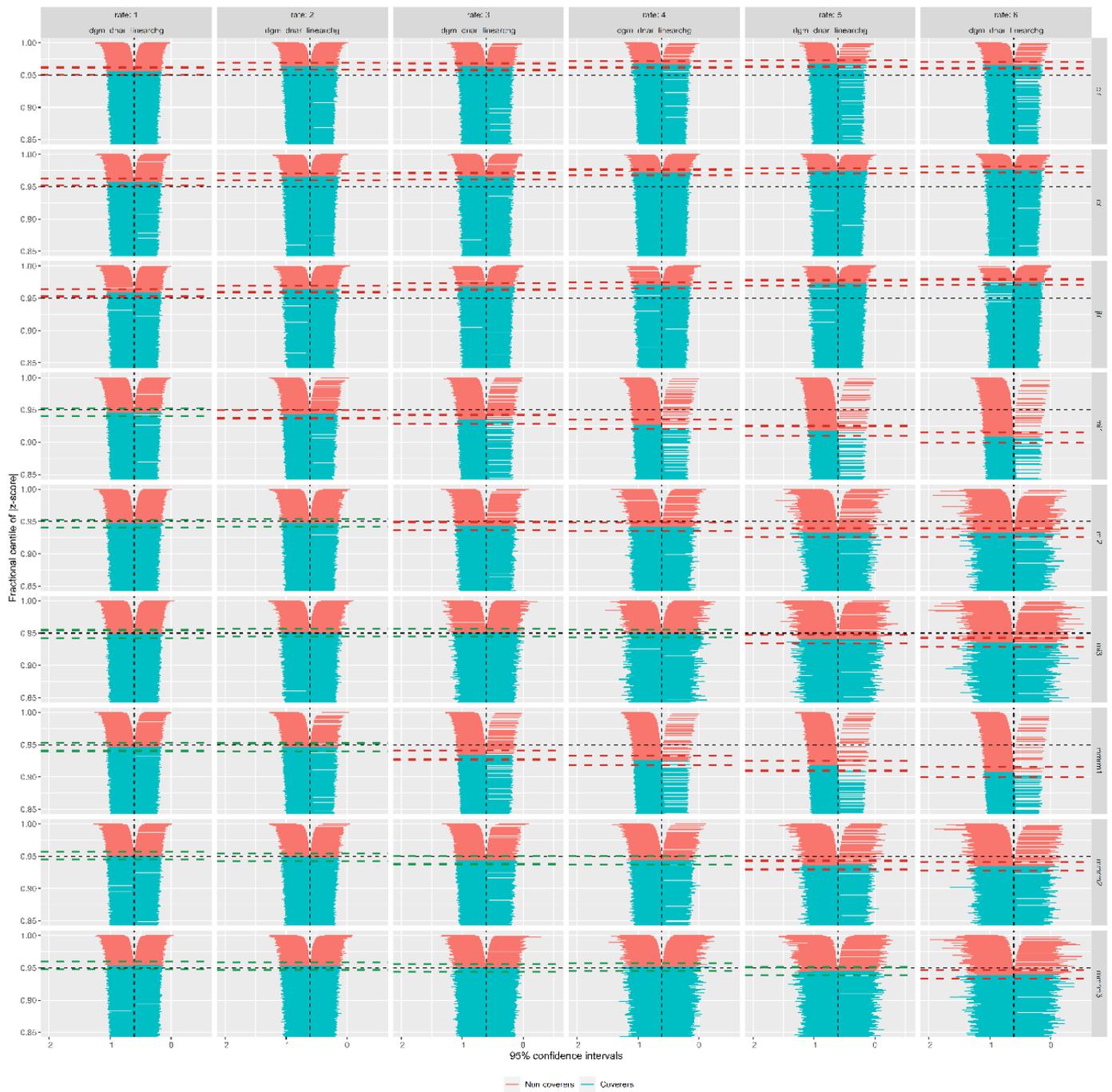

*Figure SI8.* Zipper plot for the DGM with discontinuation not at random and gradual loss of treatment effect. The zipper plot is zoomed so that only the 15% of runs with theta further away from the true theta are shown. Confidence interval for coverage is shown is green (if compatible with 95% coverage) or red (if there is under- or over-coverage)

***Figure SI9.** Zipper plot for the DGM with discontinuation at random and instantaneous loss of treatment effect. The zipper plot is zoomed so that only the 15% of runs with theta further away from the true theta are shown. Confidence interval for coverage is shown is green (if compatible with 95% coverage) or red (if there is under- or over-coverage)*

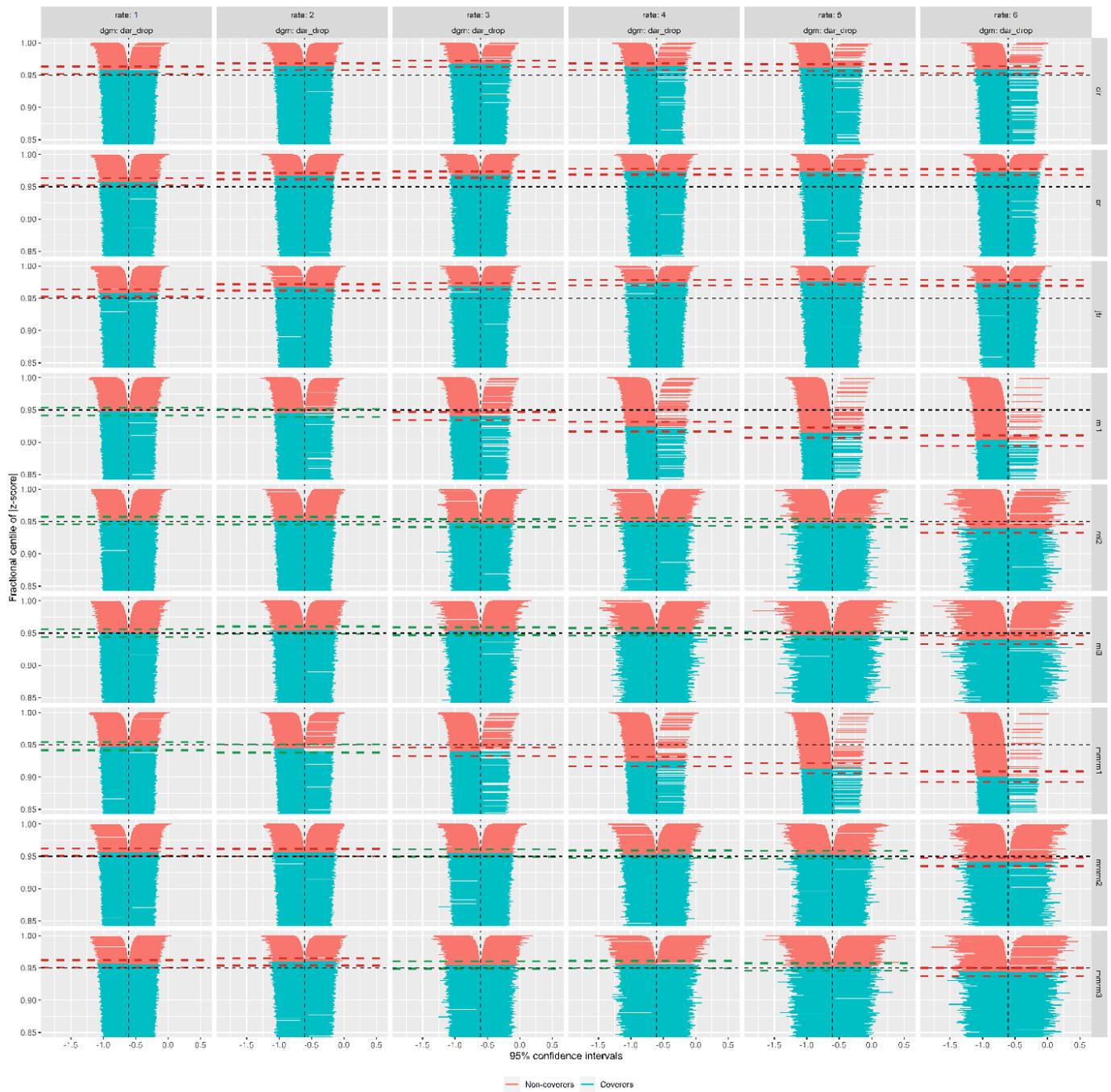

*Figure SI10.* Zipper plot for the DGM with discontinuation not at random and instantaneous loss of treatment effect. The zipper plot is zoomed so that only the 15% of runs with theta further away from the true theta are shown. Confidence interval for coverage is shown is green (if compatible with 95% coverage) or red (if there is under- or over-coverage)

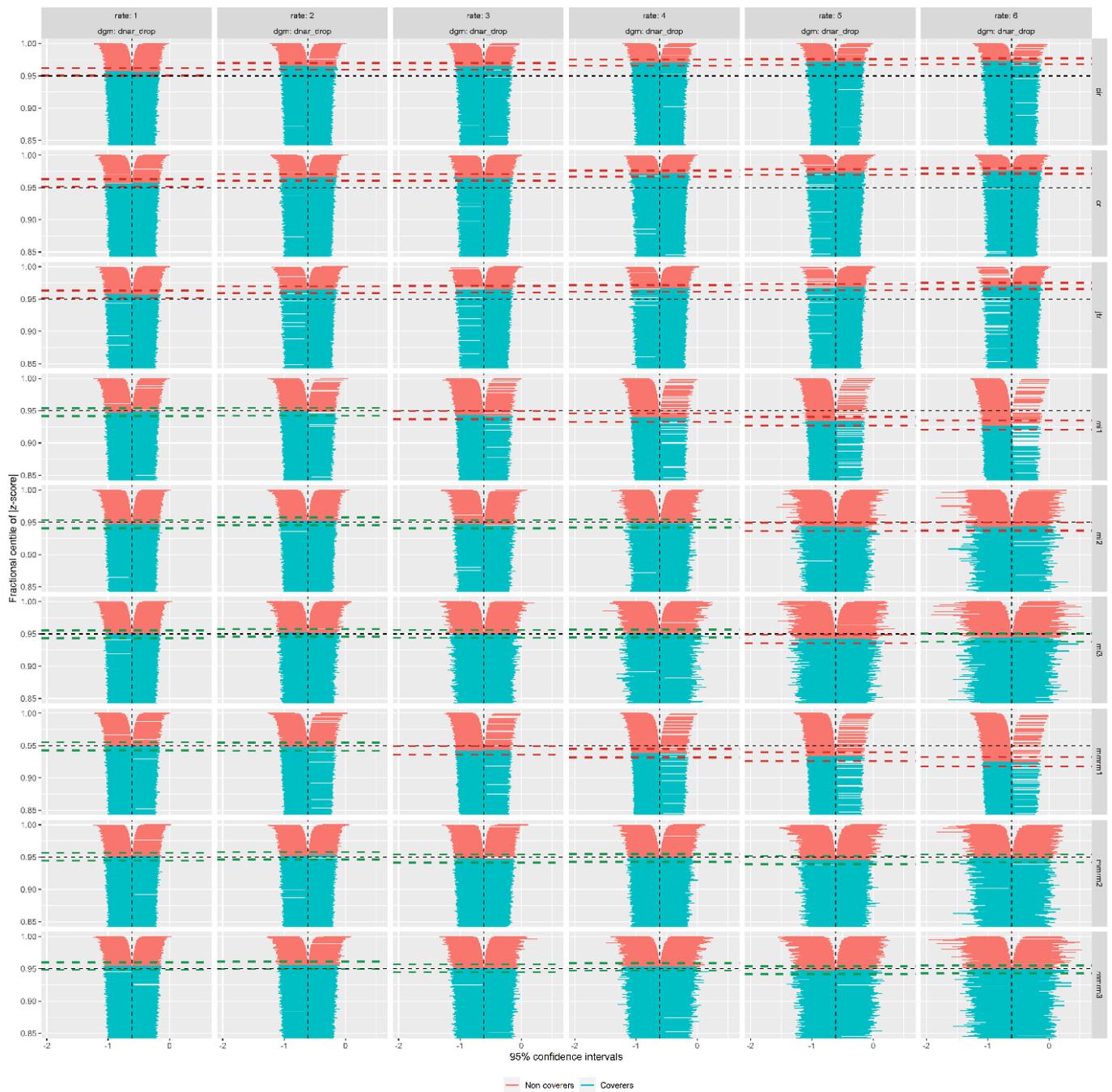